\documentclass[11pt, a4paper]{article}
\pdfoutput=1
\usepackage[utf8]{inputenc}
\usepackage[T1]{fontenc}
\usepackage{lmodern, textcomp}
\usepackage[american]{babel}
\usepackage{csquotes}
\usepackage{bbold}
\usepackage{multicol}

\usepackage{mathtools}

\usepackage{eurosym}
\usepackage
[nottoc]{tocbibind}
\usepackage{authblk}
\usepackage{fancyhdr}
\usepackage{xspace}
\usepackage{bm}

\usepackage{graphicx}
\usepackage{subcaption}
\usepackage{float}
\usepackage{makecell}
\usepackage{multirow}
\usepackage{multicol} 
\usepackage{marginnote}
\usepackage[multiple]{footmisc}

\usepackage[usenames, dvipsnames, svgnames, table]{xcolor}

\usepackage[backend=bibtex8, citestyle=numeric-comp, maxbibnames=99,
			sorting=none, sortcase=false, sortcites=true, giveninits=true]{biblatex}

\usepackage{amsmath, amsfonts, amssymb}
\usepackage{mathtools}
\usepackage{mathrsfs}
\usepackage{cancel}
\usepackage{braket}
\usepackage{tensor}
\usepackage{slashed}
\usepackage{siunitx}
\usepackage{listings}

\usepackage{stmaryrd}
\usepackage{caption}
\usepackage{relsize,exscale}
\usepackage{array}
\usepackage[toc,page]{appendix}

\usepackage{enumerate}

\DeclareSymbolFont{largesymbols}{OMX}{cmex}{m}{n}

\setcellgapes{1pt}
\makegapedcells
\newcolumntype{R}[1]{>{\raggedleft\arraybackslash }b{#1}}
\newcolumntype{L}[1]{>{\raggedright\arraybackslash }b{#1}}
\newcolumntype{C}[1]{>{\centering\arraybackslash }b{#1}}

\newcommand{\Tr}{\mathrm{Tr}}

\newtheorem{theorem}{Theorem}
\newtheorem{definition}{Definition}

\newtheorem{remark}{Remark}
\newtheorem{claim}{Claim}

\newtheorem{lemma}{Lemma}

\newcommand{\beq}{\begin{equation}}
\newcommand{\eeq}{\end{equation}}
\newcommand{\bea}{\begin{eqnarray}}
\newcommand{\eea}{\end{eqnarray}}
\definecolor{mygray}{gray}{0.3}

\newcommand{\bes}{\begin{eqnarray}}
\newcommand{\ees}{\end{eqnarray}}

\newcommand\restr[2]{{
  \left.\kern-\nulldelimiterspace 
  #1 
  \vphantom{\big|} 
  \right|_{#2} 
  }}
\def\extd{\mathrm {d}}

\newcommand{\N}{\mathrm{N}}
\newcommand{\GU}{\mathrm{GUE}}
\newcommand{\RE}{\mathrm{Re}}
\newcommand{\IM}{\mathrm{Im}}
\newcommand{\Sol}{\mathrm{Sol}}
\usepackage{multicol}
\usepackage{amssymb}
\def\Xint#1{\mathchoice
   {\XXint\displaystyle\textstyle{#1}}%
   {\XXint\textstyle\scriptstyle{#1}}%
   {\XXint\scriptstyle\scriptscriptstyle{#1}}%
   {\XXint\scriptscriptstyle\scriptscriptstyle{#1}}%
   \!\int}
\def\XXint#1#2#3{{\setbox0=\hbox{$#1{#2#3}{\int}$}
     \vcenter{\hbox{$#2#3$}}\kern-.5\wd0}}

\def\dashint{\Xint-}

\usepackage[heightrounded, top=4cm, bottom=4cm, left=3cm, right=3cm, headheight=14pt]{geometry}

\usepackage[pdftex, breaklinks=true, linktocpage,
	colorlinks=true, urlcolor=blue, linkcolor=blue, citecolor=red
]{hyperref}

\newcommand{\email}[1]{\href{mailto:#1}{\nolinkurl{#1}}}
\newcommand{\emailfoot}[1]{\thanks{\email{#1}}}

\usepackage{marginnote}
\newcounter{draftcommentcnt}
\NewDocumentCommand{\draftcomment}{s O{red} m}{%
	\def\margnote{\IfBooleanTF{#1}{\marginnote}{\marginpar}}%
	\stepcounter{draftcommentcnt}%
	\textcolor{#2}{#3}%
	\margnote{\textcolor{#2}{$\Leftarrow$ \arabic{draftcommentcnt}}}%
}

\fancypagestyle{preprint}{
	\fancyhf{}

	\fancyhead[R]{\textsf{\preprint}}
}

\numberwithin{equation}{section}

\hypersetup{
	pdftitle={Functional renormalization group for $p=2$ matrix dynamics in the planar approximation},
}

\title{Functional renormalization group for ‘‘$p=2$'' like glassy matrices in the planar approximation\\
\bigskip
\Large{I. Vertex expansion at equilibrium}}

\author[1]{Vincent Lahoche\emailfoot{vincent.lahoche@cea.fr}}
\author[1,2]{Dine Ousmane Samary\emailfoot{dine.ousmanesamary@cipma.uac.bj}}

\affil[1]{%
	Université Paris Saclay, \textsc{Cea}, \textsc{List}, Gif-sur-Yvette, F-91191, France
}
\affil[2]{%
	Faculté des Sciences et Techniques (ICMPA-UNESCO Chair)
	\protect\\
	Université d'Abomey-Calavi, 072 BP 50, Bénin
}
\begin{document}
\maketitle

\hrule
\hrule
\begin{abstract}
In this paper, we study the equilibrium states of a $\N\times \N$ stochastic complex random matrix $M$, whose entries evolve in time accordingly with a Langevin equation including both Gaussian white noises and a linear disorder, materialized by the Wigner random matrices. In large $\N$-limit, the disorders behave as effective kinetics, and we examine a coarse-graining over the Wigner spectrum accordingly with two different schemes that we call respectively ‘‘active'' and ‘‘passive''. We then investigate explicit solutions of the nonperturbative renormalization group using vertex and derivative expansion, a simple way to deal with the nonlocal nature of the effective field theory at large $\N$. Our main statement is the existence of well-behaved fixed point solutions and at least some evidence about a discontinuous (first order) phase transition between a condensed and a dilute phase. We finally interpret the resulting phase space regarding the out-of-equilibrium process related to the dynamical phase transitions. 
\end{abstract}

\hrule

\hrule
\newpage
\pdfbookmark[1]{\contentsname}{toc}
\tableofcontents
\bigskip

\newpage

\section{Introduction}

Historically, the spin-glass theory was born, as the Ising model for magnetic systems, from the idealization of the strange behavior of some magnetic alloys containing strong interacting ions (impurities) disturbing the background weakly interacting lattice \cite{cugliandolo2002dynamics}. These impurities, randomly distributed in every piece of matter, are characterized by their weak time-scale evolution compared with, for instance, magnetic dipoles involved in the magnetization of the alloys (spins). Such a disorder is said to be \textit{quenched}, and an explicit model of such a quenched interaction is for instance provided by Ruderman-Kittel-Kasuya-Yosida (RKKY) theory \cite{altieri2023introduction}. The most striking fact linked to these impurities is the frustration: It becomes impossible along a closed loop to satisfy with discrete spins all the couplings at the same time.  This phenomenon is at the origin, at low temperatures, of the appearance of a new phenomenon characteristic of glassy states (the appearance of a big number of metastable states in which the system finds itself trapped arbitrarily for a long time). 
\medskip

Note that despite the existence of explicit models, the glass transition phenomenon is very general, and even appears in situations where exact theories like RKKY do not apply. This becomes a general phenomenon occurring as soon as there is a sufficient number of impurities, distributed randomly from one piece of matter to another \cite{de2006random,nishimori2001statistical,mezard1987spin,castellani2005spin}. As an example we can be considered the Ising model for ferromagnetism; the archetypal spin glass model being the so-called Edwards-Anderson model:
\begin{equation}
H_{\text{EA}}:=-\sum_{\langle i,j \rangle}\, J_{ij} s_i s_j\,,
\end{equation}
where in this Hamiltonian, the sum is performed over
nearest-neighbor spins $s_i=\pm 1$ and the \textit{disorder} $J_{ij}$ materializing impurities is a $\N\times \N$ Gaussian random matrix \cite{nishimori2001statistical}. There are several theoretical signatures of the glass transition, depending on the type of approach considered, and we will speak of \textit{replica symmetry breaking} in so-called static approaches, \textit{weak ergodicity breaking} (dynamical transition) and \textit{aging effects} in the dynamical approach and \textit{complexity} for approaches based on Thouless-Anderson-Palmer (TAP) formalism \cite{nishimori2001statistical,mezard1987spin}. Note that all these approaches are usually considered complementary, and in particular do not predict the same value for the glassy transition temperature. 
The $p=2$ spherical spin glass model is particularly revealing of this last point \cite{de2006random}. In the static limit, it behaves as a ferromagnet in disguise, with a phase transition toward a ferromagnetic phase having a macroscopic occupation number along the disorder's eigendirection corresponding to the maximal eigenvalue. Hence, no replica symmetry breaking is expected in that case, and it is easy to check that the fully symmetric replica solution is always stable. From the dynamical point of view, however, the behavior of the system is not as trivial. It exhibits weak ergodicity breaking and aging effect \cite{cugliandolo1995full}, and fails to reach equilibrium in the low-temperature phase for generic initial conditions, a situation reminiscent of domain coarsening physics occurring in phase ordering kinetics physics \cite{bray2002theory}.
\medskip

The renormalization group, which is one of the powerful tools in theoretical physics to discuss scaling effects and effective physique \cite{Zinn-Justin:1989rgp,ZinnJustinBook2} is nevertheless rarely used in the literature of glassy systems (especially nonperturbative renormalization techniques). There are nevertheless some references that deal with this issue (see for instance \cite{pimentel2002spin,parisi2001renormalization,castellana2015hierarchical,banavar1988heisenberg,castellana2010renormalization,angelini2017real} and references therein for standard spin glass) and \cite{de2006random,tarjus2004nonperturbative,tarjus2008nonperturbative,balog2018criticality} for applications of perturbative and nonperturbative techniques for random field Ising model. Recently we have considered this field of research, particularly concerning applications of the nonperturbative functional renormalization group to the study of the long-time behavior of the dynamics of standard $p$-spin models \cite{lahoche2021functional,lahoche2023functional}, as well as \cite{erbin2023functional} proposing an application of the formalism in data science for signal detection. Let us notice that there are many reasons to include the renormalization group in this kind of problem. One of them is for instance the exceptional flexibility of formalism. It is well known that the renormalization group can be considered, for a very large kind of problem, covering almost all the area of physics \cite{dupuis2021nonperturbative}. Renormalization has indeed this exceptional ability to project ‘‘beyond" the approximations used to construct the flow itself.  In the case of  random matrices, the renormalization group in the planar sector predicts phase transition with a fixed point having characteristic matching with double scaling, despite the effects involved in principal sectors beyond the strict planar sector \cite{Lahoche:2019ocf,brezin1992renormalization}
\medskip

In this paper, we investigate \textit{nonperturbative renormalization group} techniques for a specific class of the stochastic matrices involving a $p=2$ type disorder coupling in their interactions. In this model, we limit ourselves to a matrix disorder itself, materialized by a Wigner matrix, and we will only consider the large   $\N$-limit. This choice will allow us to convert the disorder term into an effective kinetic term, whose spectrum, given by Wigner's law, will provide an intrinsic notion of scale from which we can construct a renormalization group following the model we have already developed in \cite{lahoche2023functional}. This approach, despite its large $\N$ limitation, avoids dealing with non-local interactions coming from averaging over disorder \cite{lahoche2021functional}, whether it concerns non-localities in time for the dynamic version, or non-localities in replicas for the equilibrium limit of the theory. In this first investigation \cite{lahoche2021functional}, we have focused on equilibrium states, and constructed a renormalization group for the expected equilibrium states of the dynamical system, assuming they are reached for a time large enough. This work is divided into two parts. The first part, which is the topic of this paper focuses on the derivative and vertex expansion, suitable because of the non-local nature of the underlying field theory. The second part, in preparation \cite{LahocheSamaryPrepa}, considers a method used in reference \cite{Lahoche:2018oeo} using Ward identities to close the hierarchical flow equations around the local sector. We also planned to extend this study with an out-of-equilibrium investigation in a forthcoming work, and finally, to investigate the behavior of the matrix model from analytical and numerical methods to evaluate the reliability of the nonperturbative framework we constructed. 
\medskip 

This investigation is the first of a program, aiming to construct renormalization group techniques for that kind of model and to investigate their reliability regarding some analytical and numerical results, which is also a work in progress. The interest for us is essentially that these models where encountered in our research for signal detection are nearly continuous spectra and aim to go beyond our previous works \cite{lahoche2023functionalD,lahoche2021field,lahoche2021generalized,lahoche2021signal,erbin2023functional}. Note that, this  follows our works about reliability of functional renormalization group for $p$-spin, again considered in our investigations about data spectra with blind cut-off between ‘‘noise'' and ‘‘information'' \cite{lahoche2021functional,lahoche2023functional}. 

The manuscript is organized as follows.  In section \eqref{SectionTechnical}, we define the model and recall the
basic derivation of Langevin and Fokker-Planck equation at the equilibrium states. We also provide at large $\N$ limit the so called effective kinetics and construct the basis ingredients of the functional renormalization group formalism applicable to our model.  In  section \eqref{vertexexpansion1} we construct the vertex expansion for equilibrium theory using particularly two schemes and discuss the numerical investigation. Section \eqref{Conclusion} is devoted to the conclusion of our work. We also give some appendices \eqref{App1}, \eqref{App3} and \eqref{App6} to clarify a certain details used throughout our investgation.

\section{Technical preliminaries}\label{SectionTechnical}

This section is divided into two distinct parts. In the first part corresponding to the first two subsections, we define the model, notations, and conventions that we  use in this paper. We also define the so called equilibrium states and construct the effective large $\N$ field theory, whose effective kinetics are provided by the well-known Wigner theorem. In the second part, we construct the general framework considered in this paper to investigate the renormalization group based on a coarse-graining over the asymptotic spectrum of the effective kinetic kernel. 

\subsection{The model and conventions}

Let us start by setting out the important objectives of this paper. We intend to investigate the large-time equilibrium states of a stochastic complex matrix $M: \mathbb{R} \to \mathbb{C}^{\N\times \N}$. We call time the underlying variable $t\in \mathbb{R}$ on which the entries of the random matrix are assumed to depend. The dynamics equations  of these matrices, by taking its components $M_{ij}$ are assumed to be first order in time, as usual for Dyson Brownian motion \cite{potters2020first}:
\begin{equation}
\boxed{\frac{\extd M_{ij}}{\extd t}=-\frac{\delta \mathcal{H}}{\delta \overline{M}_{ij}(t)}+B_{ij}(t)\,,}\label{eq1}
\end{equation}
where we used the notation $\overline{z}$ to specify the complex conjugate of the complex number $z\in \mathbb{C}$, and the entries $B_{ij}(t)$ are assumed to be distributed accordingly with a Gaussian distribution, with probability measure $\extd \mu(B)$:
\begin{align}
\extd \mu (B)=\frac{1}{z_B}\, \exp \left(- \frac{1}{T}\int_{-\infty}^{+\infty} \extd t\,\Tr\, \overline{B}^\tau(t) B(t)  \right)\extd B \extd\overline{B}\,,
\end{align}
where the Lebesgue measure $\extd B \extd\overline{B}$ is nothing but:
\begin{equation}
\extd B \extd\overline{B}:=\prod_{i,j} \extd \RE({B})_{ij}\extd \IM({B})_{ij}\,,
\end{equation}
and the normalization term $z_B^{-1}$ is chosen such that:
$
\int \, \extd \mu (B) = 1\,.
$
This is equivalent to the given of the $2$-points correlation function:
\begin{equation}
\langle \bar{B}_{i^\prime j^\prime}(t^\prime) B_{ij}(t) \rangle= T\delta_{ii^\prime}\delta_{jj^\prime}\delta(t-t^\prime)
\end{equation}
and in the rest of this paper the bracket notation $\langle X(B) \rangle $ denotes the noise averaging of the functional $X(B)$:
$
\langle X(B) \rangle := \int \extd \mu (B)\, X(B)\,.
$
\begin{remark}
From the thermodynamic interpretation point of view,
the parameter $T$ which is the standard deviation of the matrix $B$ is identified as the temperature of the equilibrium theory.
\end{remark}
The  Hamiltonian $\mathcal{H}$ is decompose as two contributions, 
\begin{equation}
\mathcal{H}= \mathcal{H}_0+\mathcal{H}_1\,,\label{decompositionH}
\end{equation}
where the \textit{interaction part} of the model is given by:
\begin{equation}
\mathcal{H}_0=\sum_{p=1}^\infty\,\frac{a_p \N^{-p+1}}{(p!)^2} \,\int_{-\infty}^{+\infty} \extd t \,\Tr \, ({M}^\dagger(t)M(t))^p=: \int_{-\infty}^{+\infty} \extd t \,H_0^{(\mathbb{C})}(M(t),\overline{M}(t))\,.\label{H0complex}
\end{equation}
Note that we discarded multi-trace interactions. This Hamiltonian is in particular invariant under the $(\mathcal{U}(\N))^{\times 2}$ Gauge transformation:
\begin{equation}
M(t) \to U M(t) V \,,\quad U,V\in \mathcal{U}(\N)\,,
\end{equation}
with $U$ and $V$ two \textit{independents} $\N \times \N$ unitary matrices. We recall the definition of the functional derivative in \eqref{eq1}, 
\begin{equation}
\frac{\delta M_{ij}(t^\prime)}{\delta M_{kl}(t)}=\frac{\delta \overline{M}_{ij}(t^\prime)}{\delta \overline{M}_{kl}(t)}=\delta_{ik}\delta_{jl}\delta(t-t^\prime)\,,\quad \frac{\delta \overline{M}_{ij}(t^\prime)}{\delta M_{kl}(t)}=0\,.\label{derivativecomplexrules}
\end{equation}
In some applications considered in the literature for field theories having such a kind of internal symmetry, it is usual to make use of the Ribbon graph representation, which makes explicit the strand structure of the Feynman graphs and which we will sometimes use in this paper. But we rather make use of the so-called \textit{bubble representation}, mainly considered in the random tensor and tensorial field theory literature \cite{guruau2017random}.
\medskip

\begin{figure}
\begin{center}
\includegraphics[scale=0.8]{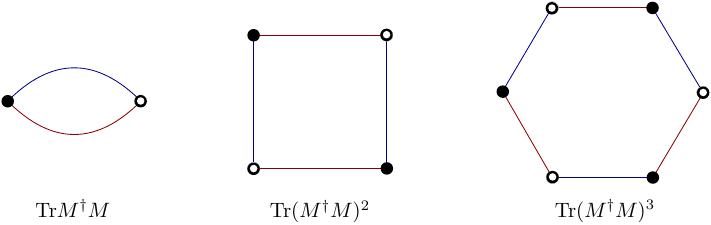}
\end{center}
\caption{Bubble graph representation for complex matrices.}\label{fig2}
\end{figure}

In the bubble representation, the interactions are materialized by a colored bipartite regular graph, such that black and white vertices materialize respectively fields $M$ and $\overline{M}$. Two colored edges, corresponding to the two indices are hooked to each node, and these edges are hooked together, according to the trace structure. For instance:
\begin{equation}
\vcenter{\hbox{\includegraphics[scale=0.7]{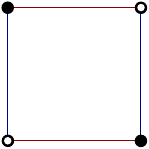}}}\,\equiv \,\sum_{{\color{blue}i},{\color{red}j},{\color{blue}k},{\color{red}l}} M_{{\color{blue}i}{\color{red}j}}\bar{M}_{{\color{blue}k}{\color{red}j}}M_{{\color{blue}k}{\color{red}l}}\bar{M}_{{\color{blue}i}{\color{red}l}}
\end{equation}
Figure \ref{fig2} provides some examples. Note that the graph has to be connected for a single trace interaction. In the examples pictured in Figure \ref{fig2} we focused on connected trace invariant, whose bubble is made of a single piece. We could also have considered interactions made of several connected parts. In this article, we will only consider connected interactions, and we will adopt the following definition:

\begin{definition}\label{deflocal}
A function of matrix field $M(t)$ at time $t$ is said to be ‘‘local" (with respect to the underlying unitary symmetry) if it takes the form of a single trace, i.e. if its power field expansion only involves connected components. 
\end{definition}
The remaining contribution $\mathcal{H}_1$ in the decomposition \eqref{decompositionH} is the \textit{disorder Hamiltonian}:
\begin{align}
\mathcal{H}_1[M,\overline{M}]&=\int_{-\infty}^{+\infty} \extd t\,\Tr \left( J M(t) M^\dagger(t) \right)+\Tr \left( K M^\dagger(t) M(t) \right)\\
&=: \int_{-\infty}^{+\infty} \extd t\, H_1^{(\mathbb{C})}[M(t),\overline{M}(t)]\,,
\end{align}
where both $J$ and $K$ are \textbf{Hermitian Wigner matrices}, for which we recall the definition and the main statement due to Wigner \cite{potters2020first,forrester2010log}:
\begin{definition}
The \textbf{Wigner ensemble} for hermitian or symmetric matrices is the set of complex Hermitian matrices, such that entries on the diagonal chosen independently from a zero mean distribution $\kappa_{\text{d}}$, and upper triangular entries chosen independently from a finite mean distribution $\kappa_{\text{od}}$ with finite variance $\sigma^2<\infty$. Lower triangular entries are defined from the hermitian (or symmetric) condition.
\end{definition}
This definition is quite general, and the main statement of random matrix theory is the \textit{Wigner theorem}:

\begin{theorem}\label{theoremWigner}
Let $A$ be a $\N\times \N$ Wigner matrix, with eigenvalues $\{\lambda_1,\lambda_2,\cdots,\lambda_\N\}$. Then, for any suitable test function $f$, we have the equality for $\N\to \infty$:
\begin{equation}
\frac{1}{\N}\lim_{\N \to \infty} \sum_{\mu=1}^\N\, f(\lambda_\mu)= \int_{-2\sigma}^{2\sigma} \, \extd \lambda \, \mu_W(\lambda) f(\lambda)\,,
\end{equation}
with:
\begin{equation}
\mu_W(\lambda)=\frac{\sqrt{4\sigma^2-\lambda^2}}{2\pi \sigma^2}\,.
\end{equation}
\end{theorem}
An elementary example of Wigner matrices is provided by the special \textit{Gaussian unitary ensemble} ($\GU$) with probability measure (Note that entries of $J$ do not depend on time):
\begin{equation}
\extd \mu(J)=\frac{1}{z_J}\, \exp \left(- \frac{1}{2\sigma^2}\, \Tr\, J^2\right)\,,\quad z_J:=\int \extd J e^{-  \frac{1}{2\sigma^2}\, \Tr\, J^2}\,.\label{defmuJ}
\end{equation}
Using the functional derivation rule \eqref{derivativecomplexrules}, we get the equations of motion in the complex case:
\begin{equation}
\dot{M}_{ij}(t)= -\sum_{k} \left(J_{ik}M_{kj}(t)+M_{ik}(t) K_{kj}\right)-\frac{\delta \mathcal{H}_0}{\delta \overline{M}_{ij}(t)} + \eta_{ij}(t)\,.
\end{equation}
It is suitable to work in the eigenbasis for disorders matrices $J$ and $K$, which we assume to have the same distribution for their entries. Because the disorder is independent of time, the projection can be done without taking into account the stochastic effects linked to \textit{Îto's theorem} \cite{potters2020first}. We denote respectively by $\{\lambda \}$ and $\{\mu\}$ the set of eigenvalues for $J$ and $K$, and by $\{{u}^{(\lambda)}_i\}$ and $\{{v}^{(\mu)}_i\}$  respectively the set of orthonormal eigenvectors, i.e. such that:
\begin{align}
\sum_{i=1}^N u_i^{(\lambda)}u_i^{(\lambda^\prime)} = \delta_{\lambda \lambda^\prime} \,,&\qquad \sum_{\lambda=1}^N u_i^{(\lambda)}u_j^{(\lambda)}= \delta_{ij}\\
\sum_{i=1}^N v_i^{(\lambda)}v_i^{(\lambda^\prime)} = \delta_{\lambda \lambda^\prime} \,,&\qquad \sum_{\lambda=1}^N v_i^{(\lambda)}v_j^{(\lambda)}= \delta_{ij}\,.
\end{align}
We furthermore define also the projected fields:
\begin{equation}
M_{\lambda \mu}(t):=\sum_{i,j} u_i^{(\lambda)} v_j^{(\mu)} M_{ij}(t)\,,\qquad \eta_{\lambda \mu}(t):=\sum_{i,j} u_i^{(\lambda)} v_j^{(\mu)} \eta_{ij}(t)\,,
\end{equation}
and the stochastic equation reads finally in the eigenspace:
\begin{equation}
\boxed{\frac{\extd}{\extd t}{M}_{\lambda\lambda^{\prime}}(t)=-(\lambda+\lambda^\prime)M_{\lambda\lambda^\prime}(t)-\frac{\partial H_0}{\partial {M}_{\lambda^\prime\lambda}(t)}+\eta_{\lambda\lambda^\prime}(t)\,.}\label{eqmove2}
\end{equation}

\subsection{Fokker-Planck equation and equilibrium states}\label{FPSec}

A stochastic equation like \eqref{eq1} admits a probabilistic interpretation.
We denote by $P(M,t)$ the probability distribution of  the random noise $\eta(t)$ which influences the trajectory such that $M(t)=M$, with the initial condition
 $M(t=0)=M_0$. Formally:
\begin{equation}
P(M,t):= \left\langle\prod_{i,j}\,  \delta(M_{ij}-M_{ij}(t)) \delta(\overline{M}_{ij}-\overline{M}_{ij}(t)) \right\rangle\,,\label{PMcomplex}
\end{equation}
The probability $P(M,t)$ follows a Fokker-Planck equation that can be easily deduced from the equation of motion \eqref{eq1}. For complex matrices, for instance, we have:
\begin{equation}
\frac{\partial P}{\partial t}= \mathbf{H} P[M,t]\,,\label{equEQ}
\end{equation}
where the Hamiltonian $\mathbf{H}$ is the second order derivative operator:
\begin{equation}
\mathbf{H}:= \sum_{i,j} \Bigg( T\frac{\partial^2}{\partial M_{ij} \partial \overline{M}_{ij}}+2\frac{\partial^2 H^{(\mathbb{C})}}{\partial M_{ij} \partial \overline{M}_{ij}}+\frac{\partial H^{(\mathbb{C})}}{\partial M_{ij}} \frac{\partial}{\partial \overline{M}_{ij}}+\frac{\partial H^{(\mathbb{C})}}{\partial \overline{M}_{ij}} \frac{\partial}{\partial {M}_{ij}}  \Bigg)\,,
\end{equation}
and equation \eqref{equEQ} admits a stationary solution $P_0^\mathbb{C}[M]$, which does not depend explicitly on time, 
\begin{equation}
P_0^\mathbb{C}[M]\propto \exp \left(- \frac{2H^{(\mathbb{C})}[M,\overline{M}]}{T} \right)\,,\label{equilibriumstate}
\end{equation}
where
\begin{equation}
H^{(\mathbb{C})}:=H^{(\mathbb{C})}_0+H^{(\mathbb{C})}_1.\label{defHC2}
\end{equation}
In the \textit{equilibrium dynamics}, the system is assumed to relax toward equilibrium theory, and:
\begin{equation}
\lim\limits_{t \rightarrow +\infty} P[M,t]\to P_0[M]\,,\label{longtime}
\end{equation}
provided that the integral converges:
\begin{equation}
\int \extd M \extd \overline{M}\, \exp \left(- \frac{2H^{(\mathbb{C})}[M,\overline{M}]}{T} \right) \leq \infty\,.\label{normalizationcondition}
\end{equation}
We recall the derivation of the Fokker-Planck equation for complex matrices in Appendix \ref{App1}, but more details about the general formalism can be found in references \cite{ZinnJustinBook2,Zinn-Justin:1989rgp}. \begin{definition}
The equilibrium generating functional $Z_{\text{eq}}^\mathbb{C}[L,\bar{L}]$ for complex matrices are respectively: 
\begin{equation}
Z_{\text{eq}}^\mathbb{C}[L,\overline{L}]:=\int \extd M \extd \overline{M}\, e^{-2 \frac{H^{(\mathbb{C})}[M,\overline{M}]}{T}+ \overline{L}\cdot M+\overline{M}\cdot L}\,,\label{parteqcomplex}
\end{equation}
where the dot product ‘‘$\cdot$'' is defined as:
$
A \cdot B:= \int \extd t\, \sum_{i,j=1}^N\, A_{ij}(t) B_{ij}(t)\,.
$
\end{definition}

\subsection{Large--$\N$ limit and effective kinetics}

Because the disorder is quenched, for equilibrium states, the construction of the averaging generally requires to use of a replica (see \cite{castellani2005spin} and references therein). However, in the large $\N$ limit, there is another route, exploiting the properties of the Wigner ensemble (Theorem \ref{theoremWigner}). Indeed, instead of integrating the disorder out, we can only consider the equations of move \eqref{eqmove2} in the large $\N$ limit, and, in the eigenbasis for them, regard disorder matrices as a non-conventional propagator. For instance, the equilibrium partition function for complex random matrices \eqref{parteqcomplex} becomes (we set $T=2$ in this section):
\begin{equation}
Z_{\text{eq}}^\mathbb{C}[L,\overline{L}]\underset{\N\to \infty}{\longrightarrow}\int \extd M \extd \overline{M}\, e^{- H_\infty^{\mathbb{C}}[M,\overline{M}]+ \overline{L}\cdot M+\overline{M}\cdot L}\,,\label{parteqcomplex2}
\end{equation}
where:
\begin{equation}
H_\infty^{\mathbb{C}}[M,\overline{M}]:= \sum_{\lambda,\mu} \overline{M}_{\lambda\mu} (\lambda+\mu+a_1){M}_{\lambda\mu}+\sum_{p=2}^\infty\,\frac{a_p \N^{-p+1}}{(p!)^2} \,\Tr \, ({M}^\dagger(t)M(t))^p\,.\label{HamiltonianCdef}
\end{equation}
Here, $\lambda$ and $\mu$ are interpreted as nearly continuous variables, and the large $\N$ limit model looks like a non-local field theory with equilibrium propagator:
\begin{equation}
C_{\lambda\mu}=\frac{1}{\lambda+\mu+a_1}\,.
\end{equation}
It is tempting to interpret the eigenvalues $\lambda,\mu$ as the analog of momenta, but they are not positive definite. For this reason, we define the positive quantities (in the strict large $\N$ limit) $p_1$ and $p_2$ as:
\begin{equation}
p_1:=\lambda+2\sigma \,, p_2:= \mu+2\sigma\,,
\end{equation}
and we define the ‘‘mass" $m$ as:
\begin{equation}
m:=a_1-4\sigma\,,
\end{equation}
such that the propagator becomes:
$
C_{p_1p_2}:=E^{-1}(p_1,p_2)\,,
$
where $E(p_1,p_2):=p_1+p_2+m$ is the \textit{energy} of the mode $(p_1,p_2)$. Note that the disorders $J$ and $K$ have the same variance $\sigma$. The resulting field theory is reminiscent of the well-known Kontsevich matrix model \cite{Francesco_1995}, and exactly the starting point followed by Cugliandolo and Dean \cite{cugliandolo1995full} and the authors in \cite{lahoche2023low} to solve the $p=2$ spherical spin glass dynamics in the large $\N$ limit. We can however expect that this effective field theory will be difficult to solve exactly because the trick consisting of closing the equation of move exploiting the self-averaging of the square length of the random vector is no longer relevant for matrices. The resulting non-conventional field theory can however be investigated using the standard perturbation theory -- see the one loop computation done in Appendix \ref{App3}.
\medskip

We will instead focus on studying the behavior of the large-scale system by the non-perturbative renormalization group, to which this article is largely dedicated. Indeed, the non-trivial scaling provided by the propagator allows us to consider a renormalization group \textit{à la Wilson} \cite{ZinnJustinBook2,wilson1983renormalization,wilson1971renormalization}, integrating out fields having large momenta to construct an effective field theory in the low energy regime. Furthermore, because of the momenta distribution:
\begin{align}
\rho(p):=\frac{\sqrt{p(4\sigma-p)}}{2\pi \sigma^2}\,,\label{WigPdis}
\end{align}
behaves like $\rho(p)\sim p^{\frac{1}{2}}$, the asymptotic field theory is expected to be close to a non-local Euclidean field theory in three dimensions \cite{lahoche2023functionalD,lahoche2023functional}. Indeed, $p_1+p_2$ plays the same role as $\vec{p}\,^2$ in ordinary field theory, but due to the non-local structure of interaction (see below), only one of the two variables is integrated along the loops involved in Feynman diagrams. Then, integrals are all the form $\int_0^{4\sigma} \rho(p) F(p)$, which have to be compared with the typical integrals $\int \extd^D\vec{p}\, \tilde{F}(\vec{p}\,)$ involved in ordinary quantum field theories. Furthermore, the Euclidean measure $\extd^D \vec{p}\,$ shows that the values of $\vec{p}\,^2$ are distributed as $\rho_{\mathbb{R}^D}(\vec{p}\,^2)\sim (\vec{p}\,^2)^{\frac{D-2}{2}}$, and matches with the asymptotic behavior of the Wigner distribution as $D=3$. In the reference \cite{lahoche2023functional} of the same authors, such a nonperturbative renormalization group has been considered for the $p=2$ spherical model, which in the eigenbasis of the disorder exhibits such generalized momentum. The relevance of such a non-conventional field theory has furthermore been pointed out again about renormalization in the series of papers \cite{lahoche2023functionalD,lahoche2021field,lahoche2021generalized}, related to the signal detection issue for nearly continuous spectra. 
\medskip

\subsection{The functional renormalization group formalism}

In this section, we introduce the functional renormalization group, based on a coarse-graining over the Wigner spectra. We consider two different schemes, which we refer to as Scheme 1 and Scheme 2 in the rest of this paper. 

\subsubsection{Coarse graining over Wigner spectrum}\label{secCG1}

\begin{figure}
\begin{center}
\includegraphics[scale=0.7]{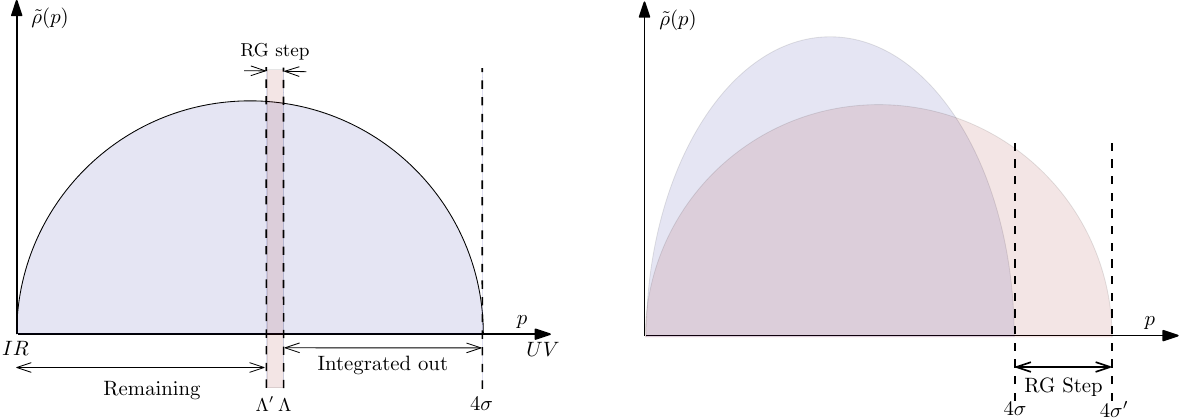}
\end{center}
\caption{Qualitative illustration of the typical renormalization group step performed on the Wigner spectra. On the left, typical windows of momenta integrated out. On the right, the typical deformation of the shape of the distribution after such a step, passing from $\sigma$ to $\sigma^\prime$}\label{figactivepassive}
\end{figure}

In analogy with standard field theory, one can view the modes $p_1$ and $p_2$ as momenta, corresponding to (Gaussian) fluctuations of different sizes $C_{p_1,p_2}$, namely:
\begin{equation}
\langle M_{p_1 p_2} \overline{M}_{p_1^\prime p_2^\prime} \rangle_{\text{G}} = \delta_{p_1p_1^\prime}\delta_{p_2 p_2^\prime}\, C_{p_1,p_2}\,,\label{freepropa}
\end{equation}
where the subscript $\text{G}$ is for ‘‘Gaussian". Therefore, the spectrum for the kinetic kernel $C^{-1}$ provides us a non-trivial notion of scale. This scale allows defining an objective notion of what is ‘‘ultraviolet" and what is ‘‘infrared", as it was summarized on the left of Figure \ref{figactivepassive} (the reason why Figure on the right will be clarified later). 

\begin{figure}
\begin{center}
\includegraphics[scale=0.5]{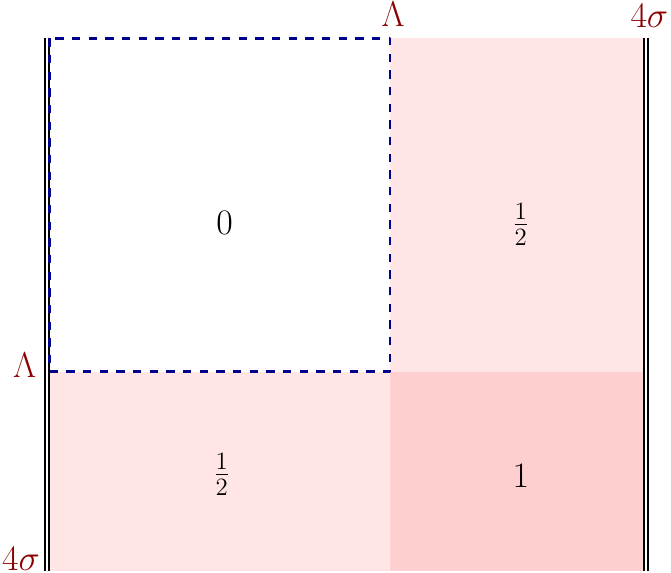}
\end{center}
\caption{The weights of integrated degrees of freedom for the RG step at the scale $\Lambda$.}\label{figpartial}
\end{figure}

We consider ‘‘UV” the degrees of freedom with large momenta, and ‘‘IR” the degrees of freedom having small momenta. This point of view allows the construction of a specific RG flow from a partial integration procedure of these degrees of freedom regarding this order relation. It has to be noticed that there is no specific rotational invariance attached to momenta $p_1$ and $p_2$, they are independent, and for this reason, the partial integration procedure is to be performed \textit{independently} on rows and columns of the matrix field $M_{p_1p_2}$, as Figure \ref{figactivepassive} shows. The corresponding RG flow can be constructed from the Wilson-Polchinski framework. Indeed, for complex matrices, for instance, we replace the full bare propagator \eqref{freepropa} by:
\begin{equation}
\langle M_{p_1 p_2} \bar{M}_{p_1^\prime p_2^\prime } \rangle_{\text{G}}^{(\Lambda)}:=  \delta_{p_1p_1^\prime}\delta_{p_2 p_2^\prime}\, D_{p_1p_2}^{(\Lambda)}\,,
\end{equation}
with:
\begin{equation}
D_{p_1p_2}^{(\Lambda)}:=\frac{1}{2}\frac{\theta(\Lambda-p_1)+\theta(\Lambda-p_2)}{p_1+p_2+m}\,,
\end{equation}
where $\theta$ is the standard Heaviside step function. We therefore attribute a weight $1$ as both $p_1$ and $p_2$ are larger than $\Lambda$, and $0$ in the opposite case. Furthermore, if only one of the momenta is larger than $\Lambda$, the loop contribution receives a weight $1/2$ as Figure \ref{figpartial} summarizes. One can therefore study the way the classical action changes as the cut-off $\Lambda$, which is continuous in the large $N$ limit, changes, what is described by the standard Polchinski equation \cite{Zinn-Justin:1989rgp}:
\begin{equation}
\boxed{\frac{\extd \mathcal S_{\text{int}}}{\extd\Lambda}=-\sum_{p_1,p_2}\frac{\extd D^{(\Lambda)}_{p_1,p_2}}{\extd\Lambda}\Big[\frac{\partial^2 \mathcal S_{\text{int}}}{\partial M_{p_1p_2}\partial \overline{M}_{p_1p_2}}-\frac{\partial \mathcal S_{\text{int}}}{\partial M_{p_1p_2}}\frac{\partial \mathcal S_{\text{int}}}{\partial \overline{M}_{p_1p_2}}\Big]\,,} \label{pol}
\end{equation}
where $S_{\text{int}}$ is the interaction part of the action, involving the power of the matrix field higher than two. Usually, before integrating out the degrees of freedom from $\Lambda$ to $e^{-\varepsilon} \Lambda$ for $\varepsilon<1$, we perform a dilatation $p\to p e^{+\varepsilon} \equiv p^\prime$, and the integration measure changes as:
\begin{equation}
\rho(p) \extd p \to \rho(e^{\varepsilon} p) \extd (e^{\varepsilon} p)=\frac{\sqrt{p^\prime(4{\sigma}^\prime-p^\prime)}}{2\pi {\sigma^\prime}^2}\, \extd p^\prime\,,
\end{equation}
where $\sigma^\prime=e^{-\epsilon} \sigma$. Hence, the shape of the generalized momenta distribution changes at each step of the renormalization group, as can be seen on the right of Figure \ref{figactivepassive}. 
\medskip

\begin{figure}
\begin{center}
\includegraphics[scale=0.5]{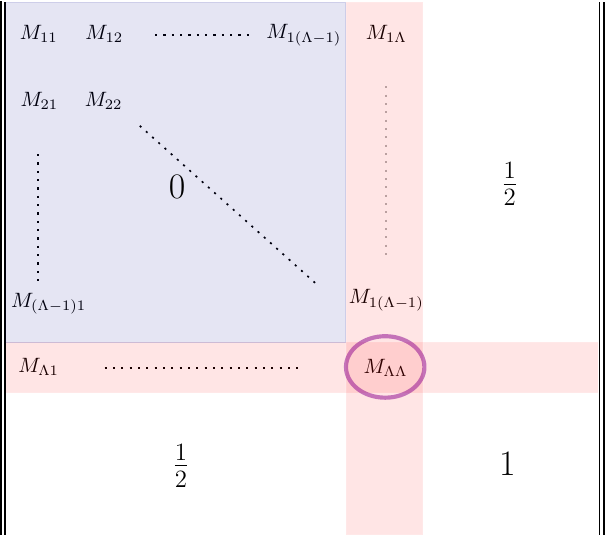} \qquad\qquad \includegraphics[scale=1]{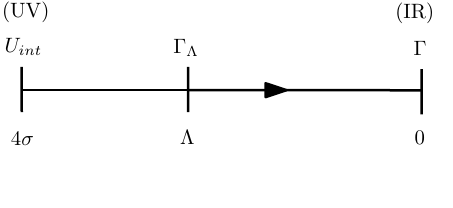}
\end{center}
\caption{The integrated degrees of freedom for the renormalization group step $\Lambda\to \Lambda-1$. $\Lambda$ (on the left). Summary of the renormalization group trajectory from UV to IR (on the right). }\label{figpartial2}
\end{figure}

Equation \eqref{pol} describes how the microscopic action, and especially the coupling constant, changes as $\Lambda$ changes. However, it should be more convenient to work with a parameter $\Lambda$ that interpolates between the initial model defined by the partition function $Z_{\text{eq}}^{(\mathbb{C})}[L,\overline{L}]$ (equation \eqref{parteqcomplex}) and some microscopic equation describing the classical limit of the theory\footnote{i.e. where quantum fluctuations are frozen.}. This could be done by the following propagator:
\begin{equation}
C_{p_1p_2}^{(\Lambda)}:=\frac{1}{2}\frac{\chi_\Lambda(p_1)+\chi_\Lambda(p_2)}{p_1+p_2+m}\,,\label{regularization1}
\end{equation}
where $\chi_\Lambda(x)$ is the windows function:
\begin{equation}
\chi_\Lambda(x):=\theta(4\sigma-x)-\theta(\Lambda-x)\,,
\end{equation}
which is equal to 1 inside the interval $(\Lambda,4\sigma)$ and 0 outside. To construct properly the large $\N$ limit into the sums, it is more convenient to understand $\theta(x)$ as the limit of a sequence of smooth functions $\theta_n(x)$, which converge weakly on the Schwartz space toward the step function $\theta(x)$ as $n\to \infty$, and we assume to take implicitly the limit after the limit $\N\to \infty$ to replace sums by integrals. For instance:
\begin{equation}
\theta_n(x):=\sqrt{\frac{n}{\pi}}\int_0^x \extd y\,e^{-n y^2}\,,\label{regultheta}
\end{equation}
holds, both for discrete or continuous\footnote{Note that this regularization is compatible with the condition $\theta(x=0)=0$. Note furthermore that the Wigner distribution vanishes on its boundaries as well.} values of $n$. This remark took into account, let us consider the partition function:
\begin{equation}
Z_{\text{eq}}^{(\mathbb{C},\Lambda)}[L]:=\int \extd M \extd\overline{M} e^{-\sum_{p_1,p_2} \overline{M}_{p_1p_2} (C^{(\Lambda)})^{-1}_{p_1p_2} M_{p_1p_2}-U_{\text{int}}[M,\overline{M}]+\overline{L}\cdot M+\overline{M}\cdot L}\,,\label{functionpartLambda}
\end{equation}
where $U_{\text{int}}[M,\overline{M}]:=\sum_{n=2}^\infty [a_n \N^{-n+1}/(n!)^2]\, \Tr (M^\dagger M)^n$. As $\Lambda \to 0$ in the integral \eqref{functionpartLambda}, we recover the initial model $Z_{\text{eq}}[L]$. In particular, the effective action:
\begin{equation}
\Gamma_\Lambda^{(0)}[\Phi]:=\overline{L}\cdot \Phi+\overline{\Phi}\cdot L-\ln Z_{\text{eq}}[L]\,,\label{effectiveaction0}
\end{equation}
where $\Phi$ and $\bar{\Phi}$ denote the classical fields, reduces to the full effective action $\Gamma$ for the initial model, as $\Lambda=0$\footnote{The reason for the index $0$ will be clear later.}. More interesting is the opposite limit, namely $\Lambda \to 4\sigma$. In this limit, the sums are empty, and the quantum fluctuations as computed in the previous section remain irrelevant from a large factor $\leq \mathcal{O}(1/\N)$, as $\N\to \infty$. Because quantum fluctuation is frozen, we expect the effective action to reduce to the classical action, which can be formally proved as follows. Because of the regularization \eqref{regultheta}:
\begin{equation}
\chi_\Lambda(x) =\epsilon \,\delta_n(4\sigma-x)+\mathcal{O}(\epsilon^2)\,,
\end{equation}
where $\epsilon:=4\sigma-\Lambda$ and:
\begin{equation}
\delta_n(x):=\sqrt{\frac{n}{\pi}} \,e^{-n (2\sigma-x)^2}\,,\label{discretedelta}
\end{equation}
is a regularized Dirac delta. Let us define: 
\begin{equation}
\Gamma_\Lambda[\Phi]+\overline{\Phi}\cdot R_\Lambda \cdot \Phi:=\overline{L}\cdot \Phi+\overline{\Phi}\cdot L-\ln Z_{\text{eq}}[L]\,,\label{GammaTrue}
\end{equation}
where:
\begin{equation}
\overline{\Phi}\cdot R_\Lambda \cdot \Phi:=\sum_{p_1,p_2} \overline{\Phi}_{p_1p_2}(R_\Lambda)_{p_2p_1} \Phi_{p_1p_2}\,,
\end{equation}
for some matrix $R_\Lambda$, whose entries will be defined below. We recall the definition of the classical field:
\begin{equation}
\Phi:= \frac{\delta }{\delta \overline{L}}\ln Z_{\text{eq}}[L]\,.
\end{equation}
By definition, we have:
\begin{equation}
e^{\overline{L}\cdot \Phi+\overline{\Phi}\cdot L-\Gamma_\Lambda[\Phi]+\overline{\Phi}\cdot R_\Lambda \cdot \Phi}= \int \extd M \extd\overline{M} e^{-\sum_{p_1,p_2} \overline{M}_{p_1p_2} (C^{(\Lambda)})^{-1}_{p_2p_1} M_{p_1p_2}-U_{\text{int}}[M,\overline{M}]+\bar{L}\cdot M+\overline{M}\cdot L}\,.
\end{equation}
The source's field $L$ and $\bar{L}$ must be replaced as:
\begin{equation}
L=\frac{\delta }{\delta \bar\Phi}\Gamma_\Lambda[\Phi]+ R_\Lambda \cdot \Phi\,,
\end{equation}
and we get:
\begin{align}
\nonumber e^{-\Gamma_\Lambda[\Phi]}&=\int \extd M \extd\overline{M} \exp\Bigg(-\overline{M} \cdot(C^{(\Lambda)})^{-1}\cdot M-U_{\text{int}}[M,\overline{M}]+\frac{\delta }{\delta \Phi}\Gamma_\Lambda\cdot (M-\Phi)\\
&+(\overline{M}-\bar{\Phi})\cdot \frac{\delta }{\delta \overline\Phi}\Gamma_\Lambda-(\overline{\Phi}-\overline{M})\cdot R_\Lambda \cdot (\Phi-M)+\overline{M}\cdot R_\Lambda \cdot M \Bigg)\,.
\end{align}
The choice $R_\Lambda=(C^{(\Lambda)})^{-1}$ cancels the first and last terms in the exponential. Furthermore, because $(C^{(\Lambda)})^{-1}\sim 1/\epsilon$ as $\Lambda\to 4\sigma$, 
\begin{equation}
\exp \left(-(\overline{\Phi}-\overline{M})\cdot R_\Lambda \cdot (\Phi-M)\right) \sim \delta(\Phi-M)\delta(\overline{\Phi}-\overline{M})\,,
\end{equation}
Hence:
\begin{equation}
\Gamma_{\Lambda \to 4\sigma}[\Phi]\to U_{\text{int}}[\Phi,\overline{\Phi}]\,.
\end{equation}
As a result, our construction interpolates between the interaction potential $U_{\text{int}}$ and the full effective action $\Gamma$, which includes all quantum corrections. The Wetterich-Morris formalism allows the transformation of this smooth interpolation as a differential equation for $\Gamma_\Lambda$. Taking the derivative of \eqref{functionpartLambda} on both sides with respect to $\Lambda$, and defining $W_\Lambda:=\ln Z_{\text{eq}}[L]$ we get:
\begin{equation}
\frac{\partial W_\Lambda}{\partial \Lambda}=-\frac{\delta}{\delta L}\cdot \frac{\partial (C^{(\Lambda)})^{-1}}{\partial \Lambda} \cdot\frac{\delta}{\delta \overline{L}} W_\Lambda+\frac{\delta W_\Lambda}{\delta L}\cdot \frac{\partial (C^{(\Lambda)})^{-1}}{\partial \Lambda}\frac{\delta W_\Lambda}{\delta \overline{L}}\,,
\end{equation}
which is equivalent to the Polchinski equation \eqref{pol}. This equation can be transformed as an equation for $\Gamma_\Lambda$ using the definition \eqref{GammaTrue} and from the observation that the previous equation assumes sources to be fixed. Replacing the derivative with respect to $\Lambda$ at fixed sources $L$ and $\bar{L}$ by a derivative with fixed classical fields $\Phi$ and $\overline{\Phi}$, 
\begin{equation}
\partial_\Lambda \vert_{L,\overline{L}}=\partial_\Lambda \vert_{\Phi,\overline{\Phi}}+ \partial_\Lambda \Phi \cdot \frac{\delta}{\delta \Phi}+ \partial_\Lambda \overline\Phi \cdot \frac{\delta}{\delta \overline\Phi}\,,
\end{equation}
we get the flow equation:
\begin{equation}
\boxed{\partial_\Lambda \Gamma_\Lambda[\Phi]=\left( \frac{\delta}{\delta L}\cdot \partial_\Lambda  (C^{(\Lambda)})^{-1} \cdot \frac{\delta}{\delta \overline{L}} \right) W_\Lambda\,.}\label{Wett}
\end{equation}
This equation takes the form of the standard Wetterich-Morris equation in the literature. Note that the second derivative concerning the sources is nothing but the $2$-point function, 
\begin{equation}
 \frac{\delta^2 W_\Lambda}{\delta L_{p_1p_2} \delta \overline{L}_{p_3p_4}}  =: (G_{\Lambda})_{p_1p_2,p_3p_3}\,,
\end{equation}
and it should be noticed from definition \eqref{GammaTrue} that:
\begin{equation}
G_{\Lambda}^{-1}:=\Gamma_\Lambda^{(2)}+(C^{(\Lambda)})^{-1}\,,
\end{equation}
where:
\begin{equation}
\Gamma_\Lambda^{(2n)}:=\frac{\delta^{2n} \Gamma_\Lambda}{\delta (\overline{\Phi})^n \delta (\Phi)^n}\,.\label{2norder}
\end{equation}
Figure \ref{figpartial2} (on the left) illustrates how degrees of freedom are integrated out from one step of the renormalization group, and Figure \ref{figpartial2} (on the right) summarizes the renormalization procedure we described above. Finally, the path we constructed from UV to IR for the equilibrium theory of complex matrices can be also considered for the corresponding MSR path integral, as we will see in the next section. Furthermore, it can also be considered for Hermitian matrices with a few modifications that we will discuss below.

\begin{remark}
In the presentation above, and the rest of this paper, we focus on the renormalization group induced by coarse-graining over the generalized momenta only. For MSR path integral however, one can also consider a coarse-graining over time (frequencies), or both in frequency and generalized momenta, as done for instance in the references \cite{duclut2017frequency, lahoche2022stochastic,lahoche2023functional}). These papers show that double coarse-graining in frequency and moment slightly increases the precision of the results, and although this is not the focus of this work, we have already planned to consider it in a future article.
\end{remark}

\subsubsection{Interpolating from classical action}\label{secCG2}
 
\begin{figure}
\begin{center}
\includegraphics[scale=1]{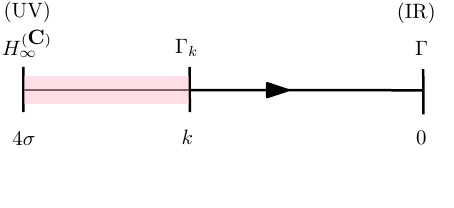}
\end{center}
\caption{The standard Wilson-Kadanoff renormalization group path for equilibrium theory. The red region corresponds to the integrated out degrees of freedom.}\label{interpolation2}
\end{figure}

The previous approach for the renormalization group is built as an explicit interpolation from $\Gamma$, the full effective action, and $U_{\text{int}}$, the initial classical interactions. Hence, if the two boundary conditions have a clear physical interpretation, the interpretation of $\Gamma_\Lambda$ is however unclear for $4\sigma > \Lambda > 0$. Indeed, the renormalization group as conceived by Wilson and Kadanoff aims to interpolate between a microscopic model and a macroscopic, effective description (see Figure \ref{interpolation2}). 

In the Wetterich formalism for instance \cite{Berges_2002,Delamotte_2012}, and focusing again on the equilibrium theory for complex matrices, we modify the classical Hamiltonian with a momenta dependent mass term $\Delta S_k$, 
\begin{equation}
H_\infty^{\mathbb{C}} \to H_\infty^{\mathbb{C}}+\Delta S_k,\label{regulatedH}
\end{equation}
depending on the running IR scale $k$, which is designed explicitly to perform such an interpolation $\Gamma_k$ from classical action $S$ for $k=4\sigma$ to the effective action $\Gamma$ for $k=0$. In that way, the perspective is, in some sense, reversed because even if $\Gamma_k$ is still interpreted as the effective action for integrated out modes, the cut-off $k$ is now viewed as an IR cut-off, such that $\Delta S_k$ suppresses degrees of freedom having momenta larger than $k$ in the loop integrals, but the UV cut-off $4\sigma$ remains fixed. The mass term $\Delta S_k$ reads for complex matrices:
\begin{equation}
\Delta S_k= \sum_{p_1,p_2} \overline{M}_{p_1p_2} R_k(p_1,p_2) M_{p_1p_2}\,.
\end{equation}
The momentum-dependent mass $R_k(p_1,p_2)$, usually called \textit{regulator} ensures that IR contributions to the effective action are suppressed \cite{Delamotte_2012}. More precisely, we require that:
\begin{equation}
R_{k=0}(p_1,p_1)=0\,,\qquad R_{k\to 4\sigma}(p_1,p_2)\to \infty\,.
\end{equation}
The first condition ensures that $\Gamma_{k\to \infty}=H_\infty^{\mathbb{C}}$ and $\Gamma_{k=0}\equiv \Gamma$, the full effective action. Generally, the definition of the regulator includes a condition, that low momenta decouple from long-distance physics, i.e. such that $R_k$ remains small as soon as $\vec{p}\,^2 > k^2$ but large enough and essentially scale independently as $\vec{p}\,^2 < k^2$, such that low energy modes are frozen out. A standard example of such a regulator is the so-called Litim regulator \cite{litim2000optimisation}:
\begin{equation}
R_k(\vec{p}\,^2):=(k^2-\vec{p}\,^2)\theta(k^2-\vec{p}\,^2)\,.
\end{equation}
The choice of this regulator and especially its dependence on the square momentum reflect the underlying rotational invariance of the theory. In the present case, we do have not such an invariance, and there is no reason to impose a condition on the sum $p_1+p_2$, but rather on each of them separately, as we have done previously in section \ref{secCG1}. For this reason, we impose to the regulator the following structure:
\begin{equation}
R_k(p_1,p_2)=Z(k)k\left(\frac{4\sigma}{4\sigma-k}\right)\left(2-\frac{p_1+p_2}{k}\right) f\left(\frac{p_1}{k}\right)f\left(\frac{p_2}{k}\right)\label{eqR}
\end{equation}
where the function $f(x)$ is inspired from the Litim's one:
\begin{equation}
f(x):=\theta(1-x)\,,
\end{equation}
and the reason $Z(k)$ is in equation \eqref{eqR} will be clear later. 
It is easy to check that $R_k(p_1,p_2)$ as defined by \eqref{eqR} satisfies all the required properties expected for a regulator\footnote{Note that in \cite{lahoche2023functional}, we made another choice, by considering instead a dilation of the moment scale. These two approaches are expected to give similar results, but the choice made here seems more interesting for numerical investigations.}. Note that, in contrast with the regularization chosen in the previous subsection (\eqref{regularization1}), $R_k$ vanishes if at least one of the two momenta is larger than $k$.
\medskip

The equation describing how the effective action $\Gamma_k$ changes as $k$ changes is the so-called \textit{Wetterich-Morris} equation; which reads for complex matrices:
\begin{equation}
\boxed{\dot{\Gamma}_k=\sum_{P=(p_1,p_2)}\,  \frac{\dot{R}_k(P)}{\Gamma^{(2)}_k(P)+R_k(P)}\,,}\label{Wett2}
\end{equation}
where 
\begin{equation}
\dot{X}:=k\frac{\extd X}{\extd k}\,,
\end{equation}
and, once again, $\Gamma^{2}_k(P)$ designates the second-order functional derivative to the classical fields (see \eqref{2norder}). Note that in the previous equation, we assumed that $\Gamma^{2}_k$ is diagonal, and $\Gamma^{2}_k(P)$ is indeed the diagonal element. Note that for hermitian matrices, the equation \eqref{Wett} has an overall factor $1/2$ on the right-hand side. Explicitly, the effective average action is defined as the (slightly modified) Legendre transform of the free energy $W_k:=\ln Z_{\text{eq},k}$ depending on $k$ through the regulator term $\delta S_k$:
\begin{equation}
\Gamma_k[\Phi]:=-W_k[L]+\overline{\Phi}\cdot L+\overline{L}\cdot \Phi - \sum_{p_1,p_2} \overline{\Phi}_{p_1p_2} R_k(p_1,p_2) \Phi_{p_1p_2}\,.
\end{equation}
Once again, the interpolation we just constructed can be also implemented for the MSR formalism. In this case, the regulator $\Delta S_k$ is of the form:
\begin{equation}
\Delta S_k=\int_{-\infty}^{+\infty} \extd \omega \sum_P i R_k(P) \left(\overline{\chi}_P(\omega) M_P(\omega)+\overline{M}_P(\omega) \chi_P(\omega)\right)\,.\label{defregularizateurOUT}
\end{equation}
Note that the regulator does not include a term of the form $\overline{\chi}_P\chi_P$, a condition required by the time-reversal symmetry -- see for instance \cite{lahoche2022stochastic}. 
\medskip

The Wetterich equations \eqref{Wett} and \eqref{Wett2} cannot be solved exactly, and require approximations. Indeed, the flow equation for effective couplings can be obtained by taking successive derivatives of the Wetterich equation. Unfortunately, this leads to an infinite and unsolvable hierarchical system of equations that we cannot solve without approximations. These approximations consist generally of projection along a reduced dimensional phase space, spanned by a small number of couplings, and are called truncation. The non-local structure of the field theory renders most of the approximations used in the literature obsolete. This is especially the case of the local potential approximation \cite{Berges_2002} outside of the symmetric phase. Hence, all the approximations we will consider in this paper focus on the symmetric phase, i.e. in the region of the phase space where we expect that the expansion of $\Gamma_k$ in the power of the classical field holds. Note that the same difficulty has been encountered in the context of tensorial and group field theories, characterized as well by their specific non-locality \cite{Lahoche:2018ggd,Lahoche:2018oeo}. In this context, the options are then quite limited, and in this paper, we will essentially consider two approaches to the problem. In the first one, the \textit{vertex expansion} is a finite expansion in terms of local vertices (see definition \ref{deflocal}). Unfortunately, this approximation is expected to have a bad behavior for non-local theory, especially because it lost a very relevant contribution coming from the momentum dependence of the effective vertex in the computation of the anomalous dimension - see for instance \cite{Lahoche:2018ggd,Lahoche_2019a,Lahoche_2020b} in the context of group field theory by the same authors. The other method that we will consider in this paper follows the strategy considered in these references and exploits the non-perturbative relations between effective observables coming from Ward identities to close the hierarchy in the local approximation.

\section{Vertex expansion for equilibrium theory}\label{vertexexpansion1}

The vertex expansion is the simpler method to solve the Wetterich equation. As explained above, it consists of expanding the effective action $\Gamma_k$ in local vertices, involving arbitrary higher powers of the classical fields. We will consider it for the two different ways to construct the renormalization group flow we considered in the previous section and to be clear, we introduce the following definition:

\begin{definition}
We refer as \textbf{scheme 1} or ‘‘passive scheme'' the coarse-graining scheme considered in section \ref{secCG1}, interpolating from microscopic interacting potential to the effective action and as \textbf{scheme 2} or ‘‘active scheme'' the coarse-graining considered in section \ref{secCG2}, interpolating from the microscopic Hamiltonian to the full effective action with dilatation of the variance $\sigma^2$. 
\end{definition}

Finally, all the computations for complex matrices will be done, and we indicate how the results have to be modified for Hermitian matrices.

\subsection{Regularization: Scheme 1}\label{subsectionscheme1}

\begin{figure}
\begin{center}
\includegraphics[scale=0.8]{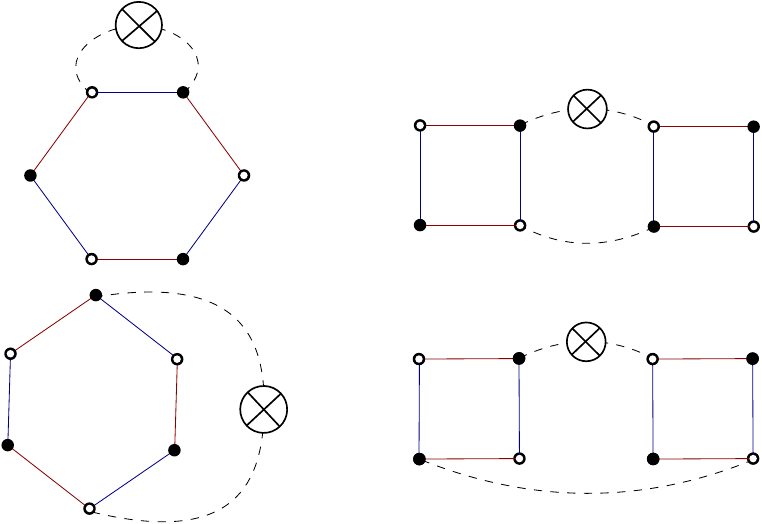}
\end{center}
\caption{Typical contributions for the flow of the quartic coupling.}\label{fig4points}
\end{figure}

\begin{figure}
\begin{center}
\includegraphics[scale=0.8]{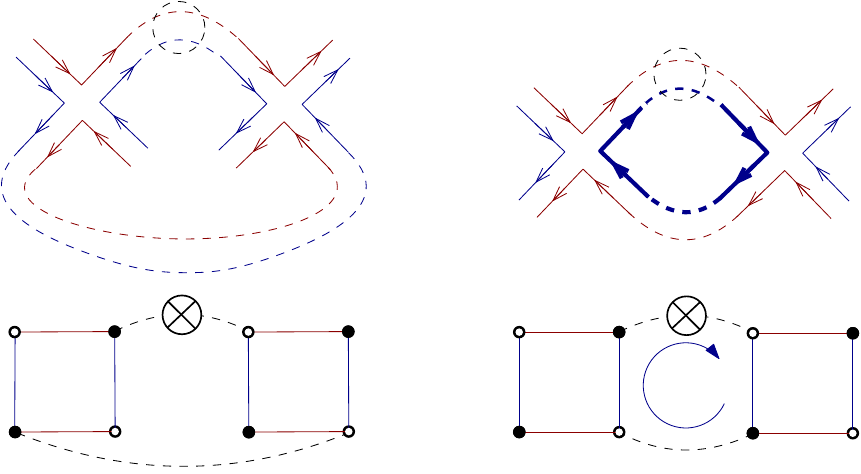}
\end{center}
\caption{Graphical representation for terms involving quartic vertices, both in ribbon and bubble representation.}\label{fig4pointsR}
\end{figure}

Let us begin by investigating the scheme 1. The vertex expansion assume that $\Gamma_\Lambda$ can be expanded as:
\begin{equation}
\Gamma_\Lambda[\Phi]=2z(\Lambda)\sum_{p_1,p_2}\bar{\Phi}_{p_1p_2} (p_1+p_2+\bar{\mu}_2(\Lambda))\Phi_{p_2p_1}+\sum_{k>1} \frac{\mu_{2k}(\Lambda)}{k!^2 \N^{k-1}} \Tr ({\Phi}^\dagger \Phi)^k\,,\label{truncation}
\end{equation}
where each monomial on the sum is local accordingly with definition \ref{deflocal}. Furthermore, from the boundary conditions for $\Gamma_\Lambda$, we must have:
\begin{equation}
z(4\sigma)=0\,.\label{initialcondz}
\end{equation}
Furthermore, because we work on the symmetric phase i.e. such that $\Phi=\bar{\Phi}=0$ are stable solutions on the shell of the move equation:
\begin{equation}
\frac{\delta \Gamma}{\delta \Phi}=0\,,
\end{equation}
the effective propagator ‘‘on-shell" in the deep IR ($\Lambda \to 0$) is diagonal with entries:
\begin{equation}
(G_{\Lambda})_{p_1p_2}\equiv \frac{1}{2}\frac{\chi_\Lambda(p_1)+\chi_\Lambda(p_2)}{z(p_1+p_2+\bar\mu_2)(\chi_\Lambda(p_1)+\chi_\Lambda(p_2))+(p_1+p_2+m)}\,.\label{effpropa}
\end{equation}
Note that the above expression remains adequate  $\forall \, (p_1,p_2)$ as soon as $n<\infty$. The derivative of the bare propagator $C^{(\Lambda)}$ on the other hand gives:
\begin{equation}
 \partial_\Lambda  (C^{(\Lambda)})^{-1}_{p_1p_2} =2(p_1+p_2+m) \frac{\delta_n(\Lambda-p_1)+ \delta_n(\Lambda-p_2)}{(\chi_\Lambda(p_1)+\chi_\Lambda(p_2))^2}\,,
\end{equation}
and it has to be noticed that, for $n$ large enough,
\begin{equation}
\chi^2_\Lambda(x)\approx \chi_\Lambda(x)\,.
\end{equation}
\paragraph{Flow equations for couplings.} Let us compute the flow equation for the “mass” $z\bar{\mu}_2\equiv \mu_2$. Because of the truncation \eqref{truncation}, we have, on-shell:
\begin{equation}
\Gamma_\Lambda^{(2)}(p_1=0,p_2=0)=2 z(\Lambda)\bar{\mu}_2\,.
\end{equation}
Hence, the flow equation for the mass can be obtained from the Wetterich equation \eqref{Wett}, taking the second derivative on both sides. Because we work on the symmetric phase, the non-vanishing contribution reads, graphically:
\begin{equation}
\partial_\Lambda \mu_2=- \, \frac{1}{2}\,\left\{\, \vcenter{\hbox{\includegraphics[scale=0.8]{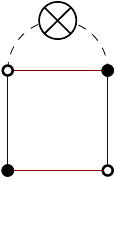}}}\,+ \, \vcenter{\hbox{\includegraphics[scale=0.8]{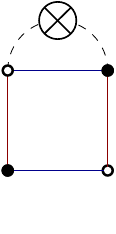}}}\,\right\}\,,\label{eqdiagmass}
\end{equation}
where the crossed disc materializes $\partial_\Lambda (C_\Lambda)^{-1}$, the dashed edges materialize the effective propagator \eqref{effpropa} and the solid edges materialize the (zero momenta) function $\Gamma_\Lambda^{(4)}$, as in the previous section. Indeed, in the previous equation, the notation is rather abstract: The quartic interaction involves $2$ fields $\Phi$ and $2$ fields $\bar{\Phi}$, materialized by black and white nodes. Solid edges are always Kronecker delta between fields, such that the trace structure of the interaction is transparent. However, free nodes have not the same interpretation as nodes hooked to a dashed edge: A dashed edge hooked between two nodes has the meaning of a Wick contraction, whereas a free black (resp. white) node is the action of the derivative operator $\delta_\Phi$ (resp. $\delta_{\bar{\Phi}}$) on the field $\Phi$ (resp. $\bar{\Phi}$) in the quartic interaction. Taking into account symmetry factors, we have the explicit expression:
\begin{equation}
\partial_\Lambda \mu_2=-\frac{\mu_4}{4\N} \sum_p (p+m)  \frac{\delta_n(\Lambda-p)}{((z+1)p+(\mu_2+m))^2}\label{flowequ2S1}
\end{equation}
where we used $\chi_\Lambda(\Lambda)=1$ as $\Lambda < 4\sigma$, because of the condition $\theta(0)=0$ compatible with our regularization scheme \eqref{regultheta}\footnote{See the remark at the end of this section.}. Because of the regularity of the function under the sum, for $n$ large enough, we can replace $p$ by $\Lambda$ everywhere. Furthermore, at fixed $n$:
\begin{equation}
\frac{1}{N}\sum_p\delta_n(\Lambda-p) \to \int_0^{4\sigma} dp \, {\rho}(p) \delta_n(\Lambda-p) \to {\rho}(\Lambda)\,.\label{sumlimit}
\end{equation}
taking $\N\to \infty$ from the first to the second step, and taking $n\to \infty$ for the final step. Hence, the flow equation reads:
\begin{equation}
\partial_\Lambda \mu_2=-\frac{1}{4}\mu_4 {\rho}(\Lambda)\frac{\Lambda+m}{((z+1)\Lambda+(\mu_2+m))^2}\,.
\end{equation}
We introduce dimensionless quantities as:
\begin{equation}
\tilde{m}:=\Lambda^{-1} m\,,\quad \tilde{\mu}_2:=(1+z)^{-1}\Lambda^{-1} ({\mu}_2+m)\,,\quad \tilde{\mu}_4:=(1+z)^{-3}\Lambda^{-1} \tilde{\rho}(\Lambda) {\mu}_4\,,\label{defdimensionless}
\end{equation}
such that the flow equation becomes:
\begin{equation}
\boxed{\Lambda\partial_\Lambda \tilde{\mu}_2=-(1+\eta)\tilde{\mu}_2-\frac{\tilde{\mu}_4}{4} \frac{1+\tilde{m}}{(1+\tilde{\mu}_2)^2}\,,}\label{flowequationmu2}
\end{equation}
where the anomalous dimension $\eta$ is defined as:
\begin{equation}
\eta:= \frac{1}{1+z} \Lambda \frac{\extd z}{\extd\Lambda}\,.
\end{equation}
In the same way, the flow equation for the $4$-point function can be computed taking the fourth derivative of the exact flow equation \eqref{Wett} on both sides. But while in the previous case, there was only one typical contribution, this time there are several contributions to take into account, all pictured in Figure \ref{fig4points}. Because we are only interested in the leading order in the large $\N$ limit, some diagrams should be discarded, and the reason why is more transparent in the Ribbon graph representation. Note that in the Ribbon representation, the propagator materialized by dotted edges in the bubble graph representation reads:
\begin{equation}
\langle M_{p_1 p_2} \bar{M}_{p_1^\prime p_2^\prime} \rangle=(G_\Lambda)_{p_1p_2}\delta_{p_1p_1^\prime}\delta_{p_2p_2^\prime}\quad \equiv \quad \vcenter{\hbox{\includegraphics[scale=1.4]{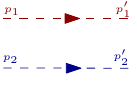}}}\,.
\end{equation}
In Figure \ref{fig4pointsR} we show the Ribbon graph representation of two of the diagrams pictured in Figure \ref{fig4points}, and it is easy to check that the main difference between them is the existence of a closed cycle on the diagram on the right. Such a closed cycle shares an additional factor $\N$ with respect to the second diagram on the left, and this diagram has to be discarded in the large $\N$ limit. We call \textit{face} such a closed cycle, and more precisely, for bubble representation:
\begin{definition}
A face is an alternate cycle made of dashed and solid edges. It can be closed or open, and closed faces, corresponding to a complete trace, share a factor $\N$.
\end{definition}\label{defface}

The length of a face is the number of dotted edges involved in the cycle, and the weight associated with a face of length $\ell$ is:
\begin{equation}
w_l(p^\prime)= \sum_p \, (G_\Lambda)_{pp^\prime}^{\ell} R_k(p,p^\prime)= \N\times \left(\int \extd p \rho(p^2) (G_\Lambda)_{pp^\prime}^{\ell} R_k(p,p^\prime)\right)\,,
\end{equation}
and share an explicit factor $\N$, the parenthesis being of order $1$. 
\medskip 

\begin{figure}
\begin{center}
\includegraphics[scale=0.8]{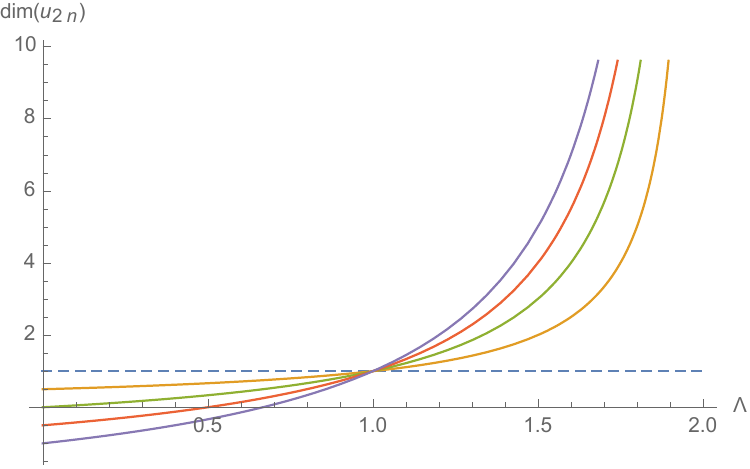}
\end{center}
\caption{Behavior of the canonical dimension $\dim(u_{2n})=\frac{2 \Lambda -\Lambda  n+n-3}{\Lambda -2}$ for $\sigma=1/2$ and $n=1$ (purple curve), $n=2$ (red curve), $n=3$ (green curve) and $n=4$ (yellow curve).}\label{figdim}
\end{figure}

The face corresponding to the closed cycle on Figure \ref{fig4pointsR} is materialized by the heavy closed blue line on the Ribbon representation, and by the blue arrow on the bubble representation. It is easy to convince ourselves that diagrams that survive at large $\N$ are those that maximize the number of created faces. These diagrams are said \textit{planar} in the literature \cite{Francesco_1995}, and their dual triangulation has zero genus. In the rest of this paper, we will use the terminology planar diagram to designate the leading order contribution to the right-hand side of flow equations. Note that because the effective diagrams provided by the Wetterich equation involve only a single loop, the number of faces is equal to $1$ or $0$. Then, keeping only the planar diagrams that create one face, the flow equation for $u_4$ reads:
\begin{align}
\nonumber \frac{1}{\N} \partial_\Lambda \mu_4=- \frac{ \mu_6}{3 \N^2} \sum_p (p+m)  \frac{\delta_n(\Lambda-p)}{((z+1)p+(\mu_2+m))^2}\\
+ \frac{\mu_4^2}{2\N^2}\sum_p (p+m)  \frac{\delta_n(\Lambda-p)}{((z+1)p+(\mu_2+m))^3}\,.
\end{align}
Hence, if we define: $\tilde{\mu}_6:=(1+z)^{-5}\Lambda^{-2} {\rho}^2(\Lambda)\mu_6$, and using the same strategy as before, we get:
\begin{equation}
\boxed{\Lambda \partial_\Lambda \tilde{\mu}_4=-(\dim(\mu_4)+3\eta)  \tilde{\mu}_4 - \frac{\tilde{\mu}_6}{3} \frac{1+\tilde{m}}{(1+\tilde{\mu}_2)^2}+ \frac{\tilde{\mu}_4^2}{2} \frac{1+\tilde{m}}{(1+\tilde{\mu}_2)^3}\,,}
\end{equation}
where the canonical dimension is defined as:
\begin{equation}
\boxed{\dim(\mu_4):= \Lambda \partial_\Lambda \ln (\Lambda {\rho}^{-1}(\Lambda))\,.}
\end{equation}
The definition of the canonical dimension can be generalized systematically, analyzing the behavior of the linear term in $\mu_{2n+2}$ involved in the flow of $\mu_{2n}$. We have, for $n>1$:
\begin{equation}
\tilde{\mu}_{2n}:=(1+z)^{-2n+1}\Lambda^{-1} ({\rho}(\Lambda))^{n-1}\mu_{2n} \,,\qquad \dim(\mu_{2n}):=\Lambda \partial_\Lambda \ln (\Lambda {(\rho}(\Lambda))^{-(n-1)})\,,\label{canonicaldimensionTrue}
\end{equation}
explicitly:
\begin{equation}
\dim(\mu_{2n}):=\frac{2 (n-3) \sigma -\Lambda  (n-2)}{\Lambda -4 \sigma }\,.
\end{equation}
Interestingly, for $\Lambda=0$, the canonical dimension reads:
\begin{equation}
\dim(\mu_{2n})\vert_{\Lambda\to 0}\rightarrow \frac{3-n}{2}\,,\label{powercountingasymptotic}
\end{equation}
which is exactly the canonical dimension of a local vertex involving $2n$ fields (measured in unities of $m^2$) for a field theory in dimension $3$, accordingly with the discussion after equation \eqref{WigPdis}. Another interesting value is reached at the middle of the spectrum, for $\Lambda=2\sigma$, for which:
\begin{equation}
\dim(\mu_{2n})\vert_{\Lambda=2\sigma}=1\,,\quad\forall\, n\,.
\end{equation}
We furthermore define the effective dimension as:
\begin{equation}
\overline{\dim}(\mu_{2n}):=\dim(\mu_{2n})+(2n-1)\eta\,.
\end{equation}
Figure \ref{figdim} shows the behavior of canonical dimensions for $\sigma=0.5$. For the sextic coupling, finally, we have:
\begin{equation}
\boxed{\Lambda \partial_\Lambda \tilde{u}_6=-\overline{\dim}(\mu_6)  \tilde{\mu}_6-\frac{\tilde{\mu}_8}{4} \frac{1+\tilde{m}}{(1+\tilde{\mu}_2)^2}+\frac{3}{2}\tilde{\mu}_4\tilde{\mu}_6\frac{1+\tilde{m}}{(1+\tilde{\mu}_2)^3}-\frac{9}{8} \tilde{\mu}_4^3 \frac{1+\tilde{m}}{(1+\tilde{\mu}_2)^4}\,.}
\end{equation}

\begin{remark}
It has to be noticed that the way the empirical distribution converges toward the Wigner distribution may introduce relevant effects in the deep IR, as soon as $\Lambda\sim 1/\N$, a domain where $1/\N$ corrections to the Wigner law could become relevant \cite{forrester2010log}. In this paper, we neglect these contributions, and we define the IR regime as $1\ll\Lambda\ll 1/\N$ (and similarly for the IR cut-off $k$ considered below).
\end{remark}

\paragraph{Anomalous dimension.} The computation of the anomalous dimension $\eta$ follows the definition of $z$ in the truncation \eqref{truncation}:
\begin{equation}
2\frac{\extd z}{\extd t}=\frac{\extd}{\extd t}\frac{\extd}{\extd p_1}\Gamma_k^{(2)}(p_1,p_2)\big\vert_{p_1=p_2=0}\,.
\end{equation}
The right-hand side can be computed from the Wetterich flow equation, as the flow equation for $\mu_2$ has been computed in the previous paragraph. Computing the derivative with respect to $p_1$ on the right-hand side of the flow equation, we have:
\begin{align}
\partial_\Lambda z=-\frac{\mu_4}{8\N}\sum_p \frac{\delta_n(\Lambda-p)}{(z(p+\bar{\mu}_2)+p+m)^2}\Bigg(1-\frac{(z+1)(p+m)}{z(p+\bar{\mu}_2)+p+m}  \Bigg)\,.
\end{align}
The sum can be computed accordingly with \eqref{sumlimit}, and introducing dimensionless quantities defined as \eqref{defdimensionless}, we get:
\begin{equation}
\boxed{\eta=-\frac{\tilde{\mu}_4}{8}\frac{\tilde{\mu}_2-\tilde{m}}{(1+\tilde{\mu}_2)^3}\,.}
\end{equation}
and because of the initial condition \eqref{initialcondz}, we have for $z(\Lambda)$, 
\begin{equation}
z(\Lambda):=\exp \left(-\int_\Lambda^{4\sigma} \eta(\Lambda)d\Lambda \right)-1\,.
\end{equation}

\subsection{Numerical investigations}

In this section, we provide comments on the numerical results obtained from the vertex expansion described in the next section. To begin, let us comment on the behavior of the canonical dimension of local vertices, shown in Figure \ref{figdim}. The canonical dimension allows understanding what is relevant and irrelevant in the vicinity of the Gaussian fixed point. Furthermore, relevant interactions usually play an important role, because for large scale, the flow is essentially driven by this quantity and the influence of irrelevant couplings is undetectable -- see \cite{Delamotte_2012,bagnuls2001exact} and reference therein. This means in particular that effective physics lost the traces of microscopic physics for large enough scales. However, in our case, because the spectrum is compact, degrees of freedom can always be integrated out from a finite number of renormalization group steps, provided that each step has a finite size. Hence, the selection procedure usually called the ‘‘large river effect'' is never complete. In the deep IR (i.e. around $\Lambda=0$), only a few numbers of couplings are relevant (which is the general case in ordinary field theories), respectively the mass, the quartic, and the sextic coupling, but higher couplings are irrelevant, as it is the case usually in standard applications of the vertex expansion. Furthermore, inverting the directions of arrows and going upstream toward the deep UV ($\Lambda \to 4\sigma$), all the couplings become relevant at the critical scale $\Lambda=2\sigma$, and their relative velocity decreases with the size of the vertex. Near the deep UV, the canonical dimensions become arbitrarily large, so that, if we start in the deep IR with an effective sextic theory not so far from the Gaussian fixed point, this theory is expected to begin arbitrarily close to the Gaussian fixed point is the deep UV. Hence, we might want to track non-Gaussian fixed points as would be the case if we were dealing with an ordinary field theory; unfortunately in this case, the canonical dimensions depend on the scale, and no global non-Gaussian fixed points can exist. As we will see, however, there are still fixed structures and attractive trajectories.
\\

Note that, rigorously speaking, such a procedure which consists of upstream the renormalization group does not make sense physically, because of its semigroup structure: the partial integration procedure is irreversible. However, mathematically, it makes sense in the parametrization we have chosen for the theory space. Then, we will assume this hypothesis holds for our numerical investigations to reverse the arrow directions. It has to be noticed that this procedure is not exceptional in the renormalization group literature. All the literature about asymptotic safety scenario \cite{eichhorn2018status,niedermaier2006asymptotic} has the same point of view, and focuses on a well-parametrized region of the theory space to investigate the UV behavior of the theory. This was also the point of view of the recent contributions exploiting the renormalization group in the context of neural network physics \cite{erbin2022non,erbin2022renormalization} and for the already cited bibliographic line addressing the signal detection issue for nearly continuous data spectra \cite{lahoche2023functionalD,erbin2023functional}. 
\medskip

Before coming up with the results, we can again make an additional observation. Even if the notion of fixed point does not make sense globally, it should be relevant at the tail of the spectra, in the deep IR, for $k$ small enough. Indeed, in this limit, 
\begin{equation}
\rho(p) = \frac{(p)^{\frac{1}{2}}}{\pi \sigma^{\frac{3}{2}}} + \mathcal{O}(p)\,,
\end{equation}
and as we explained before, the theory behaves as an ordinary (but non-local) field theory in dimension 3 (see for instance \eqref{powercountingasymptotic}). This in particular allows us to introduce the concept of \textit{asymptotic IR fixed point}:

\begin{definition}
We call the asymptotic IR fixed point, a fixed point of the asymptotic three-dimensional field theory. 
\end{definition}

In the absence of true global fixed points, we will be especially interested first in the existence of such a structure, and we will then study their persistence in the UV regime. First, consider the case $\tilde{m}=0$, and then investigate the case $\tilde{m}\neq 0$. Not that because we are interested first in the IR regime, $\vert\tilde{m}\vert$ is assumed to be a large number (of order $1/\Lambda$), which is the general case in the absence of fine-tuning of the value of $m$. 

\begin{figure}
\begin{center}
\includegraphics[scale=0.45]{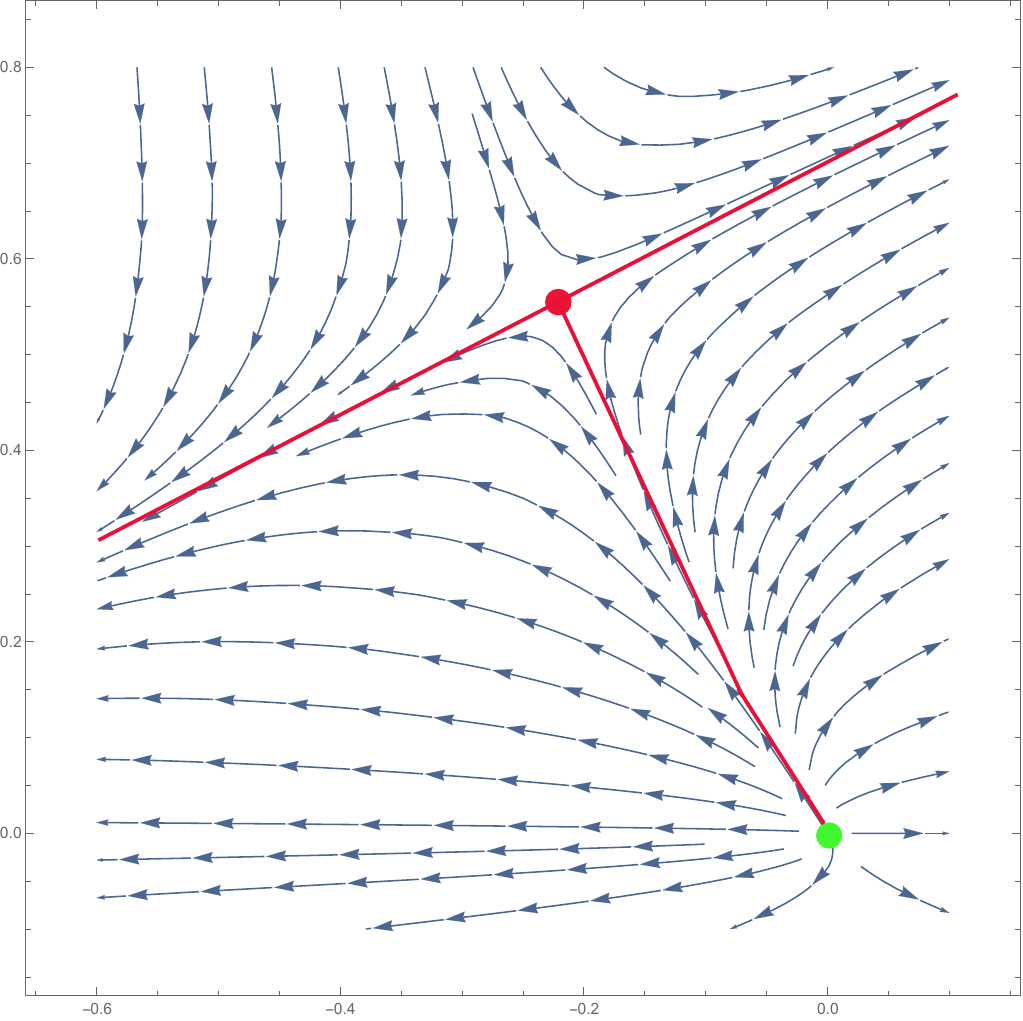}
\end{center}
\caption{Behavior of the renormalization group flow in the vicinity of the fixed points, reminiscent of a second order phase transition. The green dot is the Gaussian fixed point and the red dot is the Wilson-Fisher one.}\label{figPlotStream}
\end{figure}

\paragraph{Case $\tilde{m}= 0$.} A fixed point appears, at the lowest order for the quartic truncation, for the value: ($\sigma=1/2$):
\begin{equation}
\tilde{\mu}_2=\frac{1}{5} \left(\sqrt{15}-5\right)\approx -0.23\,,\qquad \tilde{u}_4=\frac{6}{55} \left(9-\sqrt{15}\right) \approx 0.56\,,
\end{equation}
with anomalous dimension:
\begin{equation}
\eta_*^{(1,4)}= \frac{1}{22} \left(2 \sqrt{15}-7\right)\approx 0.03\,.
\end{equation}
We call this fixed point $\text{FP1}_0^{(4)}$; it looks like a Wilson-Fisher fixed point (i.e. a second order phase transition), with critical exponents\footnote{We recall that critical exponents are the opposite of the eigenvalues of the stability matrix with entries $\partial_j \beta_i$, computed at the fixed point.}:
\begin{equation}
\Theta_{(1,4)}:=\{-0.77,0.76\}\,,
\end{equation}
meaning that it has one relevant and one irrelevant direction. Figure \ref{figPlotStream} shows the behavior of the renormalization group flow between the two fixed points in the quartic truncation.

We then have to investigate the higher order truncation to show if the scenario is reliable or not (note that the anomalous dimension is small enough, what is generally expected when the derivative expansion holds -- see \cite{Balog:2019rrg}). Unfortunately, no fixed points are recovered at order $6$ and $8$, and the two fixed points found at order $10$ do not have the characteristic of the fixed point found at the quartic order. The reliability of the second-order phase scenario discovered before is then not very reliable, at least using the vertex expansion. We thus claim that:

\begin{claim}
No reliable IR fixed point exists asymptotically for $\tilde{m}=0$ using standard vertex expansion.
\end{claim}

Now the question is, what happens in the deep UV? Are there trajectories in the theory space at which the $\beta$-functions vanish? We introduce the definition:

\begin{definition}
A ``{\textbf {fixed trajectory}}'' on the theory space is a stable trajectory along which $\beta$-functions vanish.
\end{definition}

\begin{figure}
\begin{center}
\includegraphics[scale=0.75]{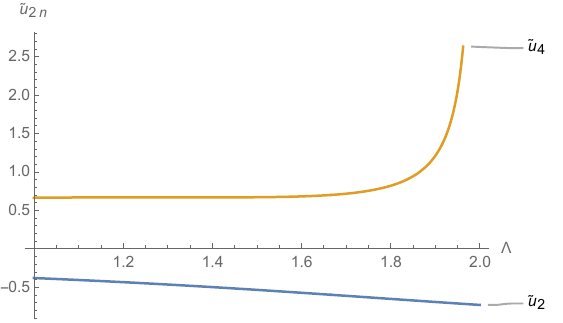}\,\includegraphics[scale=0.75]{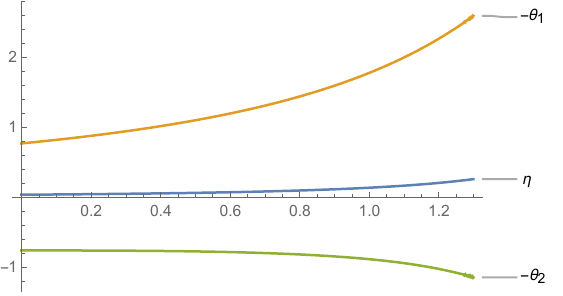}
\end{center}
\caption{On the left: Behavior of the couplings along the fixed point line in the quartic truncation. On the right: Behavior of the (opposite of) critical exponents and of the anomalous dimension.}\label{fig55}
\end{figure}

One can indeed solve the equation for each value of $\Lambda$. For the quartic order truncation, this gives the solution:

\begin{align}
\tilde{\mu}_2^*(\Lambda)&:=\frac{-\sqrt{4 \Lambda ^2-14 \Lambda +15}-2 \Lambda +5}{3 \Lambda -5}\,,\\
\tilde{\mu}_4^*(\Lambda)&:=\frac{4 \left(\sqrt{2 \Lambda  (2 \Lambda -7)+15}-\Lambda \right)^3 \left(2 \Lambda +\sqrt{2 \Lambda  (2 \Lambda -7)+15}-5\right)}{(5-3 \Lambda )^2 (\Lambda -2) \left(\Lambda +5 \left(\sqrt{2 \Lambda  (2 \Lambda -7)+15}-2\right)\right)}\,.
\end{align}
which are both plotted on the Figure \ref{fig55} (on left). Note that there is another fixed line solution, which has $\tilde{\mu}_2(\Lambda)<-1$ i.e. is disconnected from the Gaussian region due to the truncation singularity occurring for $\tilde{\mu}_2=-1$. Furthermore, the solution stops at $\Lambda=1$ (at the middle of the spectra), and at this point it becomes complex. It should be noticed again that at the limit, the value for the mass reaches $\tilde{\mu}_2=-1$, and this pathology can be interpreted as a breakdown of the vertex expansion in the symmetric phase. 
\medskip

On the right of Figure \ref{fig55} we also showed the behavior of critical exponents computed along the line, as well as the value of the anomalous dimension $\eta$. Because of their values, the two local eigen-directions are irrelevant and the fixed point line ‘‘catches'' the trajectories in its vicinity. To conclude, it should be pointed out that the anomalous dimension is positive and not too large. Unfortunately, these global trajectories disappear, as the Wilson-Fisher fixed point for higher truncations, and we can have reasonable doubt about the fact that such a fixed line is more than an additional artifact of the truncation. Despite these red flags, we conjecture that:

\begin{claim}
No reliable global fixed trajectory exists for quartic and sextic local truncations for $\tilde{m}=0$. 
\end{claim}

\paragraph{Case $\tilde{m}\neq 0$.} Now, we move on to the more complicated case, where the initial mass $m$ does not vanish. As in the previous paragraph, we begin by investigating the asymptotic IR fixed points. Since we are interested in the IR limit, the parameter $\vert\tilde{m}\vert$ is assumed to be very large, and we will first take the limit $\vert\tilde{m}\vert\to \infty$, assuming that the couplings are numbers of order $1$ ($\tilde{\mu}_{2n} \leq  \mathcal{O}(1)$). Imposing $\Lambda \partial_\Lambda{\tilde{\mu}}_2=0$, we get:
\begin{equation}
\tilde{\mu}_4^*(\tilde{m})=\frac{8 \tilde{\mu}_2 (\tilde{\mu}_2+1)^3}{-3 \tilde{m} \tilde{\mu}_2-2 \tilde{m}+\tilde{\mu}_2^2-2 \tilde{\mu}_2-2}\,.
\end{equation}
For a quartic truncation, inserting the previous solution on the explicit equation for $\Lambda \partial_\Lambda{\tilde{\mu}}_4$, find one relevant fixed point solution i.e. such that $\tilde{\mu}_2\geq -1$ (above the singularity line $\tilde{\mu}_2=0$) and different from $\tilde{\mu}_2=0$ (Gaussian fixed point). It is reached for the value:
\begin{equation}
\tilde{\mu}_2^*:=-\frac{2}{5}\,,\qquad \tilde{\mu}_4^*\approx +\frac{ 0.864}{\vert\tilde{m}\vert}\,,
\end{equation}
and this fixed point that we call $FP1_\infty^{(4)}$ looks like a Wilson-Fisher fixed point with one relevant and one irrelevant direction:
\begin{equation}
\Theta_{1,\infty}^{(4)}\approx\{0.5,\,-0.5\}\,, 
\end{equation}
and anomalous dimension
\begin{equation}
\eta_\infty^{(1,4)}=0.5\,.
\end{equation}
\medskip

This quartic fixed point solution however does not survive at order $6$, and we get a novel solution for the values:
\begin{equation}
\tilde{\mu}_2^*:=\frac{14}{13}\,,\qquad \tilde{\mu}_4^*\approx -\frac{14.76}{\vert\tilde{m}\vert}\,, \qquad \tilde{\mu}_6^*\approx \frac{134.79}{\tilde{m}^2}\,,
\end{equation}
and we call this fixed point $\text{FP1}_{\infty}^{(6)}$. Once again, this fixed point has a good sign for the last coupling, regarding the stability of the theory. Furthermore, anomalous dimension and critical exponents are given by:
\begin{equation}
\Theta_{1,\infty}^{(6)}\approx\{1.94,\, 1.44,\, 0.91\}\,, \qquad \eta_\infty^{(1,6)}\approx -0.21\,.
\end{equation}
and all the eigen-directions are relevant. This fixed point agrees with physical requirements, and we have to investigate larger truncation to investigate its reliability. At order $8$, we get essentially two relevant fixed points, namely $\text{FP1}_{\infty}^{(8)}$ and $\text{FP2}_{\infty}^{(8)}$, with critical exponents and anomalous dimension:
\begin{align}
\Theta_{1,\infty}^{(8)}\approx\{1.17,\, 0.60,\, 0.35,\, -0.04\}\,,\qquad \eta_{\infty}^{(1,8)}\approx -0.05\,,\\
\Theta_{2,\infty}^{(8)}\approx\{-2.89,\, -1.86,\, 0.54,\, -0.03\}\,,\qquad \eta_{\infty}^{(2,8)}\approx 0.27\,.
\end{align}
The first fixed point is reminiscent of the fixed point obtained at the sixth order. Also, the values of the critical exponents and the anomalous dimension are almost the same, except for a new irrelevant direction. The second fixed point is a novelty, whose stability regarding the higher order truncations should be investigated. The first fixed points are reached for the values:
\begin{equation}
\tilde{\mu}_2^*:=\frac{2}{567} \left(2 \sqrt{865}-25\right)\,,\qquad \tilde{\mu}_4^*\approx -\frac{1.20}{\vert\tilde{m}\vert}\,, \qquad \tilde{\mu}_6^*\approx \frac{3.12}{\tilde{m}^2}\,,\qquad\tilde{\mu}_8^*\approx -\frac{5.52}{\vert\tilde{m}\vert^3}\,.
\end{equation}
At order ten, we find only one relevant fixed point $\text{FP1}_{\infty}^{(10)}$ with critical exponents and anomalous dimension:
\begin{equation}
\Theta_{1,\infty}^{(10)}\approx \{{2.36,\, 1.91,\,1.88,\, 1.85,\, 1.02}\}\,,\qquad \eta_{\infty}^{(1,10)}\approx -0.26\,,
\end{equation}
and at order twelve, in the same way, we get again a fixed point $\text{FP1}_{\infty}^{(12)}$, having three relevant directions:
\begin{equation}
\Theta_{1,\infty}^{(12)}\approx \{+1.15,\, -0.90,\, +0.59,\, -0.49,\, +0.32,\, -0.08\}\,,\qquad \eta_{\infty}^{(1,12)}\approx -0.05\,.
\end{equation}
The last fixed point is again reminiscent of the fixed point $\text{FP1}_{\infty}^{(6)}$ discovered at the order $6$ and recovered at order $8$. Then, it is tempting to conjecture that:

\begin{claim}
It must exist a true interacting fixed point in the deep IR, having three relevant directions and reaching the point of the $\tilde{\mu}_2$ axis as $\tilde{m}\to \infty$. 
\end{claim}

Note that the anomalous dimension decreases with the order of the truncation, seeming to indicate that effects beyond the derivative expansion, and at the same order, these terms considered in the truncation could play a relevant role. Furthermore, because of the remaining dependency over $\tilde{m}$, the corresponding fixed point can also be viewed as a \textit{fixed trajectory}. However, the reliability of the fixed point remains poor. For instance, it is not recovered at order $10$, seeming to indicate that a powerful method than vertex expansion has to be considered, and will be the subject of a companion paper.
\medskip

In the previous analysis, we assumed $\tilde{u}_{2n} \leq  \mathcal{O}(1) \, \forall \,n$. Now we consider another limit, with the definitions:
\begin{equation}
\tilde{g}_{2n}:= \tilde{\mu}_{2n}(1+\tilde{m})^{n-1}\,,\forall n\,.
\end{equation}
from which we deduce:
\begin{equation}
\Lambda \partial_\Lambda\tilde{g}_{2n}=(1+\tilde{m})^{n-1} \Lambda \partial_\Lambda\tilde{\mu}_{2n}+(n-1)\tilde{g}_{2n} \frac{\Lambda \partial_\Lambda \tilde{m}}{1+\tilde{m}}\,,
\end{equation}
or in the deep IR, 
\begin{equation}
\Lambda \partial_\Lambda\tilde{g}_{2n}=(1+\tilde{m})^{n-1} \Lambda \partial_\Lambda\tilde{\mu}_{2n}-(n-1)\tilde{g}_{2n} \,.
\end{equation}
Hence, in the IR limit, the power counting for $\tilde{g}_{2n}$ is:
\begin{equation}
\dim(\tilde{g}_{2n})=\dim(\tilde{\mu}_{2n})+(n-1)\equiv \frac{1+n}{2}\,.
\end{equation}
Furthermore, 
\begin{equation}
\eta=-\frac{\tilde{g}_4}{8}\frac{1}{(1+\tilde{\mu}_2)^3} \frac{\tilde{\mu}_2-\tilde{m}}{1+\tilde{m}} \to \frac{\tilde{g}_4}{8}\frac{1}{(1+\tilde{\mu}_2)^3} \,.
\end{equation}
Let us investigate the resulting flow equations. At the quartic order, we find a fixed point with two irrelevant directions, for the values:
\begin{equation}
\text{FP1}_{\mathcal{\infty}}^{(4)}=\{\tilde{\mu}_2=-\frac{6}{11},\, \tilde{g}_{4}\approx 1.13\}\,,
\end{equation}
and critical exponents:
\begin{equation}
\Theta_{1,\infty}^{(4)}=\{3.1,\,-2.7\}\,,\qquad \eta_\infty^{(1,4)}=1.5\,.
\end{equation}
This fixed point is a Wilson-Fisher-like fixed point and is moreover physically relevant (especially, the quartic coupling has the ‘‘good'' sign), and as previously, we will investigate higher orders. At order $6$, we get only one relevant fixed point, say $\text{FP1}_{\mathcal{\infty}}^{(8)}$, with mass respectively given by
\begin{equation}
\tilde{u}_{2*}^{(6)}:=\frac{1}{95} \left(-61+\sqrt{1441}\right)\,.
\end{equation}
and critical exponents:
\begin{align}
\Theta_{1,\infty}^{(6)}&=\{2.95,\, -1.69,\, 1.14\}\,,\qquad \eta_\infty^{(1,6)}=0.19\,.
\end{align}
This fixed point has negative sextic couplings ($\tilde{u}_{6*}^{(6)}\simeq -1.49 $, and they have again the wrong sign regarding the stability requirement. Nevertheless, the critical exponents are the value of the anomalous dimension that remains in agreement with the fixed point discovered in the quartic order. \\

\begin{figure}
\begin{center}
\includegraphics[scale=1]{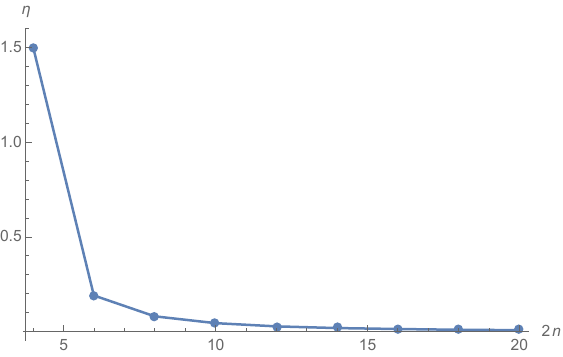}
\end{center}
\caption{Evolution of the anomalous dimension with the order of the truncation, until order $20$.}\label{figcriticalsummary}
\end{figure}

\begin{figure}
\begin{center}
\includegraphics[scale=0.7]{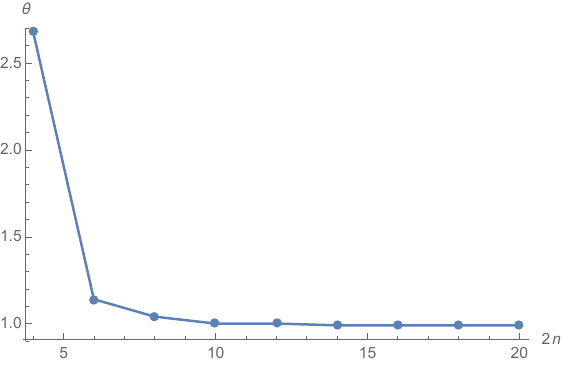}\quad \includegraphics[scale=0.7]{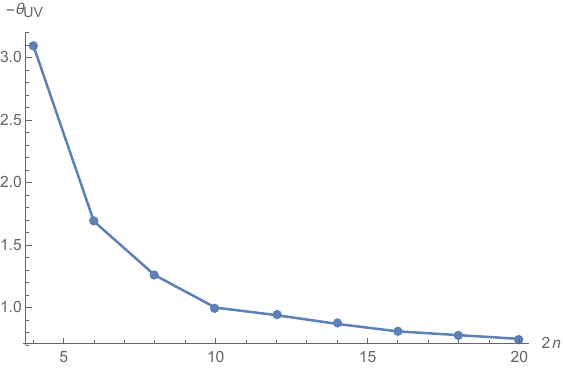}
\end{center}
\caption{Evolution of the smaller relevant critical exponent and the single irrelevant one with the order of the truncation, until order $20$.}\label{figcriticalsummary2}
\end{figure}

This fixed point seems to survive for higher orders, and its properties can be followed order by order, as summarized in  Figures \ref{figcriticalsummary} and \ref{figcriticalsummary2}. As we can see in these Figures, the fixed point seems to have good convergence properties, and both critical exponents and anomalous dimension rapidly converge with the order of the truncation, with however difficulty that the stability is not guaranteed, because the sign of the larger coupling changes with the order of the truncation, and could become very large. Note that the number of relevant directions increases with the order of the truncation, but the number of irrelevant directions remains fixed to one, and the evolution with the order of the truncation of the corresponding critical exponent is shown on the left of Figure \ref{figcriticalsummary2}. Hence, despite the previous red flags, we claim that:

\begin{claim}
It must exist an asymptotic IR fixed point of sextic order, with a single critical direction toward UV scales. \label{claimUVFP1}
\end{claim}


\subsection{Regularization: Scheme 2}\label{seceqscheme2}

In this section, we consider the second point of view (Scheme 2). We will consider separately two regimes:
\begin{enumerate}
    \item The IR regime, where it is suitable to use the approximation $\rho(p)=(\pi^{-1} \sigma^{-3/2}) \sqrt{p}$, and we will fix the normalization as $\sigma=1$.
    \item The full regime, considering the spectrum globally. 
\end{enumerate}
We will provide some details in the first case, indicating the main changes for the second case. 

\paragraph{IR Regime.} In the deep IR regime i.e. for $k$ small enough, the regulator \eqref{eqR} simplifies as:
\begin{equation}
R_k(p_1,p_2)\approx Z(k)k \left(2-\frac{p_1+p_2}{k}\right) f\left(\frac{p_1}{k}\right)f\left(\frac{p_2}{k}\right)\,,\label{eqR2}
\end{equation}
and we consider the following parametrization of the theory space:
\begin{equation}
\Gamma_k [\overline{M},M]= Z(k) \sum_{p_1,p_2} \overline{\Phi}_{p_1p_2}(p_1+p_2+2u_2(k)) \Phi_{p_1p_2}+ \sum_{n>1} \frac{u_{2n}(k)}{(n!)^2 \N^{n-1}} \Tr (\Phi^\dagger \Phi)^n\,.\label{truncationPhi2}
\end{equation}
As was the case within scheme 1 in the previous section, the non-local structure of the interaction is responsible for a non-trivial flow of the anomalous dimension, even in the symmetric phase. The computation of the flow of the mass involves the same diagram as in the previous section (equation \eqref{eqdiagmass}). More precisely, the flow equation for the mass can be obtained by taking the second derivative concerning the classical field of the Wetterich equation \eqref{Wett2}:
\begin{align}
\dot{\Gamma}_{k\,p_1p_2}^{(2)}&=\quad\vcenter{\hbox{\includegraphics[scale=0.6]{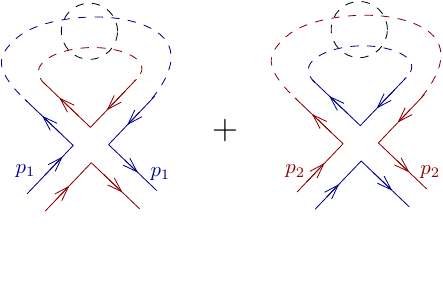}}}\\
&=-\frac{u_{4}}{8\N} \sum_{q} \frac{\dot{R}_k(q,p_1)+\dot{R}_k(p_2,q)}{(k+u_2)^2}\,.
\end{align}
where once again we assumed the $2$-point function to be diagonal, as expected in the symmetric phase \cite{Lahoche:2018oeo}. Setting the external momenta $p_1$ and $p_2$ equal to $0$, and because of the definition of $u_2$ in the truncation \eqref{truncationPhi2}, we have:
\begin{equation}
{\Gamma}_{k\,00}^{(2)}=2Z(k) u_2(k)\,,
\end{equation}
and:
\begin{align}
Z(k)(\dot{u}_2+\nu u_2)=-\frac{u_{4}Z^{-2}(k)}{8\N} \sum_{q} \frac{\dot{R}_k(q,0)}{(k+u_2)^2}\,.
\end{align}
The remaining sum over $q$ can be computed in the continuum limit in the IR approximation:
\begin{equation}
Z(k)(\dot{u}_2+\nu u_2) =-\frac{u_{4}}{8\pi} \frac{Z^{-2}(k)}{(k+u_2)^2} \int_0^4 \extd q \sqrt{q} \dot{R}_k(q,0)\,,
\end{equation}
where we defined:
\begin{equation}
\nu:=k\frac{\extd }{\extd k} \ln Z(k)\,. 
\end{equation}
The derivative of the regulator can be easily computed in the deep IR regime, 
\begin{align}
\nonumber\dot{R}_k(p_1,p_2)&=\nu {R}_k(p_1,p_2) + 2kZ(k) \theta(k-p_1)\theta(k-p_2)\\
&+Z(k)k[(k-p_1)\theta(k-p_1)\delta(k-p_2)+(k-p_2)\theta(k-p_2)\delta(k-p_1)]\label{derivativeR}\,.
\end{align}
The integral can then be computed easily, and we define:
\begin{align}
\int_0^4 \extd q \sqrt{q} \dot{R}_k(q,0)=:Z(k)k^{\frac{5}{2}} I_\nu\,,
\end{align}
with:
\begin{equation}
I_\nu(k) \equiv\frac{14\nu(k)}{15}+\frac{7}{3}\,,
\end{equation}
The flow equation for mass then becomes:
\begin{equation}
\dot{u}_2+\nu u_2 =-\frac{u_{4} k^{5/2}}{8\pi} \frac{Z^{-2}(k)I_\nu}{(k+u_2)^2} \,.
\end{equation}
At this stage, it is suitable to make use of dimensionless quantities, 
\begin{equation}
\bar{u}_2=k^{-1} u_2\,,\qquad \bar{u}_4 := Z^{-2}(k)k^{-1/2} u_4\,,\label{rescalingIR}
\end{equation}
such that the explicit dependency on $k$ of the right-hand side disappears:
\begin{equation}
\boxed{\dot{\bar{u}}_2=-(1+\nu) \bar{u}_2-\frac{\bar{u}_{4}}{8\pi} \frac{I_\nu(k)}{(1+u_2)^2}\,.}\label{beta1Scaling1}
\end{equation}
The transformations \eqref{rescalingIR} are exactly what we expect for a field theory in dimension $3$, and it is easy to check recursively that the correct rescaling for higher couplings, such that higher order equations become autonomous is:
\begin{equation}
\bar{u}_{2n}=Z^{-n}(k) k^{-\frac{3-n}{2}} u_{2n} \,.
\end{equation}
Indeed, in the same way, we get for the quartic and sextic couplings:
\begin{equation}
\boxed{\dot{\bar{u}}_4=-\left(\frac{1}{2}+2\nu\right)\bar{u}_4- \frac{\bar{u}_6}{6\pi} \frac{I_\nu(k)}{(1+\bar{u}_2)^2}+ \frac{\bar{u}_4^2}{4\pi} \frac{I_\nu(k)}{(1+\bar{u}_2)^3}\,,}\label{beta1Scaling2}
\end{equation}
and:
\begin{equation}
\boxed{\dot{\bar{u}}_6=-3\nu \bar{u}_6 - \frac{\bar{u}_8}{8\pi} \frac{I_\nu(k)}{(1+\bar{u}_2)^2} + \frac{3\bar{u}_4 \bar{u}_6}{4\pi} \frac{I_\nu(k)}{(1+\bar{u}_2)^3}-\frac{9\bar{u}_4^3}{16\pi} \frac{I_\nu(k)}{(1+\bar{u}_2)^4}\,.}
\end{equation}
To conclude, let us compute the anomalous dimension $\nu(k)$. Once again, because of the truncation \eqref{truncationPhi2}, we have:
\begin{equation}
\frac{\extd }{\extd p_1} \Gamma_{k,p_10}^{(2)}\Big\vert_{p_1=0}=Z(k)\,.
\end{equation}
Then:
\begin{equation}
k \frac{\extd Z}{\extd k} = \frac{\extd }{\extd p_1}\vcenter{\hbox{\includegraphics[scale=0.6]{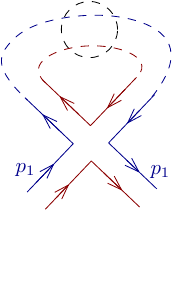}}}\,\,\Bigg\vert_{p_1=0}=-\frac{u_{4}}{8\N} \frac{\extd }{\extd p_1}\sum_{q} \frac{\dot{R}_k(q,p_1)}{(k+u_2)^2}\Big\vert_{p_1=0}\,.\label{eqflowZ}
\end{equation}
The derivative with respect to $p_1$ of $\dot{R}_k(q,p_1)$ can be easily obtained from \eqref{derivativeR}, and setting $p_1=0$, we get:
\begin{equation}
\frac{\extd }{\extd p_1} \dot{R}_k(q,0)=-\nu Z(k) \theta(k-q)-Z(k) k \delta(k-q)\,.
\end{equation}
The remaining sums can be easily computed in the continuum limit, and we get finally, because of the definition of $\nu$:
\begin{equation}
Z(k)\nu= \frac{u_{4}}{8\pi} \frac{Z(k) k^{3/2}}{(k+u_2)^2}\left(\frac{2\nu}{3}+1\right)\,,
\end{equation}
which can be solved, using dimensionless couplings, as:
\begin{equation}
\boxed{\nu(k)=\frac{1}{8\pi} \frac{\bar{u}_{4}}{(1+\bar{u}_2)^2-\frac{\bar{u}_{4}}{12\pi} }\,.}
\end{equation}
\begin{remark}
To compute $\nu$, as to compute $\eta$ in the previous section, we completely ignored the intrinsic dependency of the effective vertex on the external momenta, in agreement with the usual definition of the local potential approximation. We will see in the next part of this paper that taking into account this dependency change significantly the previous result, as it was already noticed by the same authors for tensorial field theories \cite{Lahoche:2018oeo}. 
\end{remark}

\begin{remark}
The previous equation for $\nu$ exhibits a singularity line, with an equation:
\begin{equation}
\bar{u}_{2,s}(\bar{u}_4):=\frac{1}{2} \left(\frac{\sqrt{\bar{u}_4 }}{\sqrt{3 \pi }}-2\right)\,.
\end{equation}
Hence, the physical region, connected to the Gaussian fixed point is bounded by this singularity, at which the approximation we used to solve the flow equation break-down, and fixed point solutions outside this region will be considered irrelevant. Note that in the scheme $1$, there is no such an additional singularity line, and the flow is only singular along the line $\tilde{\mu}_2=-1$.
\end{remark}

\paragraph{Full regime.} Now let us consider the general case, far from the IR regime. In this case, it is suitable to change the definition of the truncation \eqref{truncationPhi2} as:
\begin{equation}
\Gamma_k [\overline{M},M]=\frac{4Z(k)\sigma}{4\sigma-k} \sum_{p_1,p_2} \overline{\Phi}_{p_1p_2}(p_1+p_2+2u_2(k)) \Phi_{p_1p_2}+ \sum_{n>1} \frac{u_{2n}(k)}{(n!)^2\N^{n-1}} \Tr (\Phi^\dagger \Phi)^n\,,\label{truncationPhi22}
\end{equation}
such that it is suitable to define $\tilde{Z}$ as
\begin{equation}
\tilde{Z}(k):= Z(k) \, \frac{4\sigma}{4\sigma-k}\,,
\end{equation}
and
\begin{equation}
\tilde{\nu}:= k \frac{\extd }{\extd k} \ln \tilde{Z}(k)\,,
\end{equation}
in such a way that the derivative $\dot{R}_k$ of the regulator reads as \eqref{derivativeR}, replacing $\nu \to \tilde{\nu}$ and $Z \to \tilde{Z}$. The form of the flow equation for mass becomes:
\begin{align}
\tilde{Z}(k)(\dot{u}_2+\tilde{\nu} u_2)&=-\frac{u_{4}\tilde{Z}^{-2}(k)}{8\N} \sum_{q} \frac{\dot{R}_k(q,0)}{(k+u_2)^2}\\
&=-\frac{u_{4}\tilde{Z}^{-2}(k)}{8(k+u_2)^2} \int_0^{4\sigma} \rho(q)  \dot{R}_k(q,0)\,\extd q\,.
\end{align}
The remaining integral reads explicitly as:
\begin{align}
\int_0^{4\sigma}\extd q\, \rho(q)  \dot{R}_k(q,0)=\tilde{Z}\int_0^{k}\extd q \rho(q) \left[\tilde{\nu} (2k-q)+2k\right]+\tilde{Z} k \rho(k)\,.
\end{align}
Accordingly with the discussion in \cite{lahoche2022functional}, define the dimensionless quantity $\tilde{\sigma}:=k^{-1}\sigma$, such that the integral becomes:
\begin{equation}
\int_0^{4\sigma}\extd q\, \rho(q)  \dot{R}_k(q,0)=:\frac{1}{\pi} \tilde{Z}(k) k J_{\tilde{\nu}}(k)\label{definitionJ}
\end{equation}
with:
\begin{equation}
J_{\tilde{\nu}}(k)=\frac{1}{2\tilde{\sigma}^2} \left[\sqrt{4\tilde{\sigma}-1}+\int_0^1 \extd q \sqrt{q(4\tilde{\sigma}-q)}\left(\tilde{\nu}(2-q)+2\right)\right]\,.\label{defJ}
\end{equation}
The flow equation then becomes:
\begin{equation}
\dot{u}_2+\tilde{\nu} u_2=-\frac{u_{4}\tilde{Z}^{-2}(k)}{8\pi(k+u_2)^2} k J_{\tilde{\nu}}(k)\,.
\end{equation}
As for the previous case, we get rid of the explicit dependency over $k$ by a suitable redefinition of couplings:
\begin{equation}
u_2=: k \tilde{u}_2\,,\qquad u_4=:\tilde{Z}^2(k) k^2 \tilde{u}_4\,,\label{scaling2}
\end{equation}
such that the dimensionless flow equation reads:
\begin{equation}
\boxed{\dot{\tilde{u}}_2=-(1+\tilde{\nu})\tilde{u}_2-\frac{\tilde{u}_{4}}{8\pi} \frac{J_{\tilde{\nu}}(k)}{(1+\tilde{u}_2)^2}\,.}\label{equationflowu2S2}
\end{equation}
Note that the scaling relations \eqref{scaling2} differs from what we obtained in the previous paragraph (\eqref{rescalingIR}). Here, we have indeed exactly the scaling corresponding to the dimensional analysis from MSR action (see the next section), and indeed we get higher couplings:
\begin{equation}
u_{2n}=: \tilde{Z}^{n} k^n \tilde{u}_{2n}\,.\label{scalinlaw2}
\end{equation}
The explicit expressions for higher couplings can be obtained in the same way, replacing $I_\nu \to J_{\tilde{\nu}}$, $\nu \to \tilde{\nu}$ and the canonical dimension by its new value, for instance:
\begin{equation}
\boxed{\dot{\tilde{u}}_4=-2\left(1+\nu\right)\tilde{u}_4- \frac{\tilde{u}_6}{6\pi} \frac{J_{\tilde{\nu}}}{(1+\tilde{u}_2)^2}+ \frac{\tilde{u}_4^2}{4\pi} \frac{J_{\tilde{\nu}}}{(1+\tilde{u}_2)^3}\,.}\label{flowequationu4tilde}
\end{equation}
The anomalous dimension $\tilde{\nu}$ can also be computed returning on the expression \eqref{eqflowZ}, and defining the dimensionless integral:
\begin{equation}
K:=\frac{1}{2\tilde{\sigma}^2}\int_0^1 \extd q \sqrt{q(4\tilde{\sigma}-q)}=\frac{1}{2\tilde{\sigma}^2}\left(4 \tilde{\sigma}^2 \csc^{-1}\left(2 \sqrt{\tilde{\sigma}}\right)+\frac{1}{2} (1-2 \tilde{\sigma} ) \sqrt{4 \tilde{\sigma} -1}\right)\,,
\end{equation}
we have:
\begin{equation}
\boxed{\tilde{\nu}=\frac{\sqrt{4\tilde{\sigma}-1}}{16\pi \tilde{\sigma}^{2}}\frac{\tilde{u}_4}{(1+\tilde{u}_2)^2-\frac{K \tilde{u}_4}{8\pi}}\,.}
\end{equation}

\subsection{Numerical investigations}\label{NumScheme2}

We will investigate numerically the meaning of the previous flow equations. As in the previous section, we first investigate the IR regime, and we will track the fixed point in this limit first, considering the two different scalings introduced in the previous section, to which we refer as \textbf{scaling 1} and \textbf{scaling 2} respectively (note the scaling 1 is intrinsically IR).

\paragraph{Deep IR, scaling 1.} For a quartic truncation, we find only two relevant fixed points, for the values:
\begin{align}
\text{Fp1}_{41}&=\{\bar{u}_2\approx -0.21,\,\bar{u}_4 \approx 1.62\}\,,\label{fixedpointsolC1}
\end{align}
with critical exponents and anomalous dimension:
\begin{align}
\Theta_{141}&=\{-0.96,\, +0.87\}\,,\qquad \, \nu_{141}\approx 0.12\,.\\
\end{align}
and:
\begin{align}
\text{Fp2}_{41}&=\{\bar{u}_2\approx -0.58,\,\bar{u}_4 \approx 13.69\}\,,
\end{align}
with characteristics
\begin{align}
\Theta_{241}&=\{-50.47,\, +2.50\}\,,\qquad \, \nu_{241}\approx -2.87\,.\\
\end{align}
The first fixed point has the characteristics of a Wilson-Fisher fixed point, and its reliability should be evaluated for higher truncation. However, because of its characteristics, it is tempting to view the second fixed point as an irrelevant spurious artifact of the truncation. Indeed, a large but negative anomalous dimension invalidates the derivative expansion in the regularization we considered. This is in particular because of the definition of the regulator, to recover the suitable boundary condition $\Gamma_{k\to \infty}=H_\infty^{\mathbb{C}}$, the regulator must behave as $k^r$ with $r>0$ for large $k$. In the vicinity of an interacting fixed point furthermore, with anomalous dimension $\eta_*$, $Z=k^{\eta_*}$, giving the prefactor $k^{1+\eta_*}$, then:

\begin{lemma}
Using the Limit regulator \eqref{eqR2}, the anomalous dimension has to satisfy the following self-consistent bound:
\begin{equation}
\boxed{\eta_* > -1}\,.\label{boundeta}
\end{equation}
\end{lemma}
This condition is violated by the fixed point solution $\text{Fp1}_{41}$, and the quartic sector is empty. 
\medskip

\begin{figure}
\begin{center}
\includegraphics[scale=0.5]{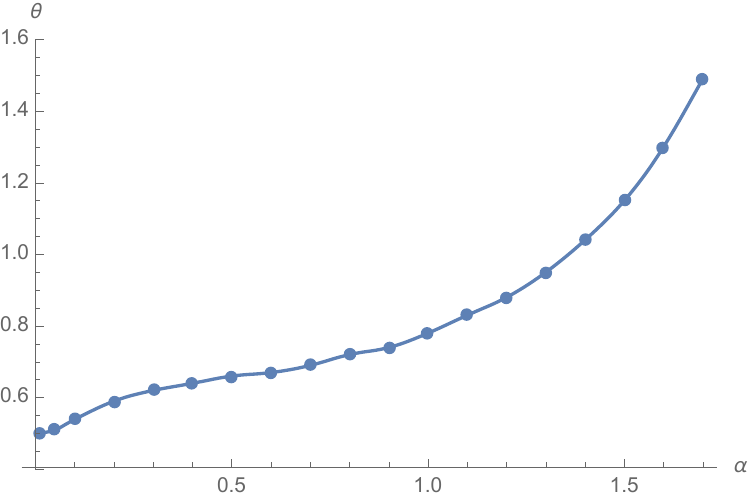}\quad\includegraphics[scale=0.5]{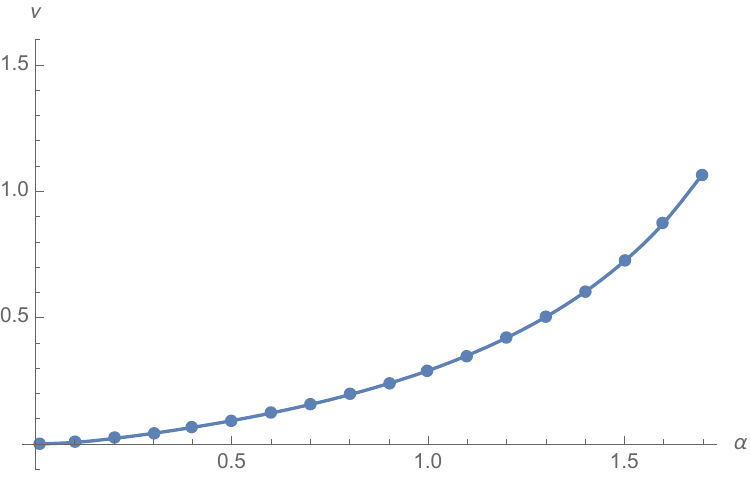}
\end{center}
\caption{On the left: dependency on $\alpha$ of the relevant critical exponent $\theta$. On the right: dependency on $\alpha$ of the anomalous dimension.}\label{depalpha}
\end{figure}

\begin{figure}
\begin{center}
\includegraphics[scale=0.35]{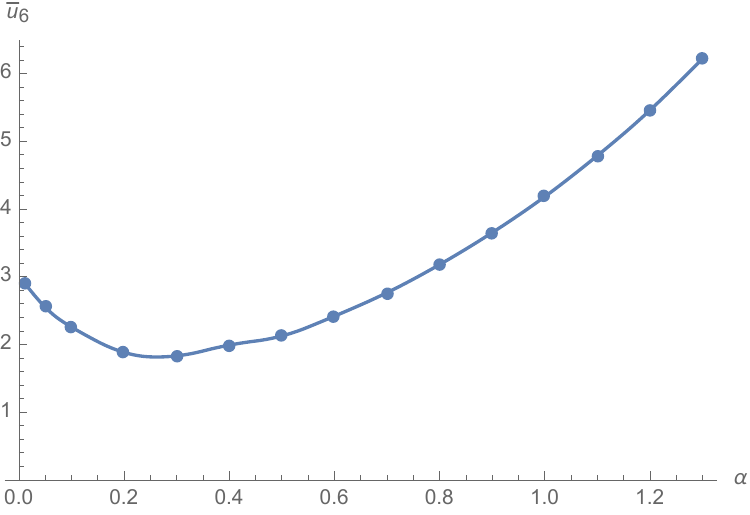}\quad\includegraphics[scale=0.35]{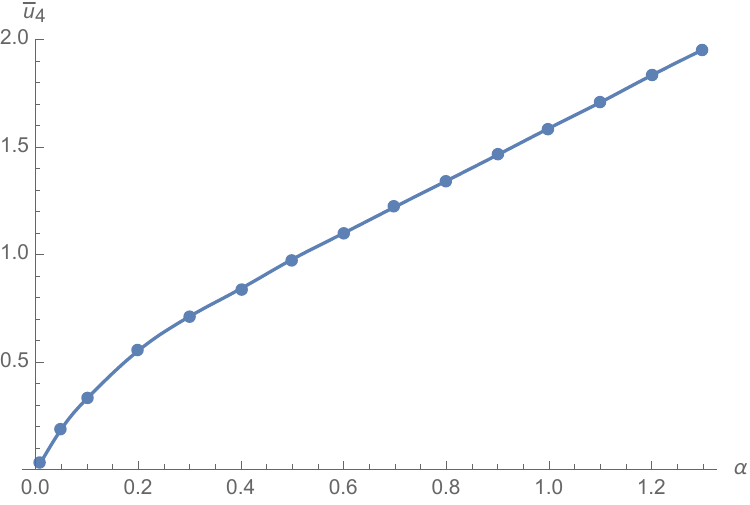}\quad \includegraphics[scale=0.35]{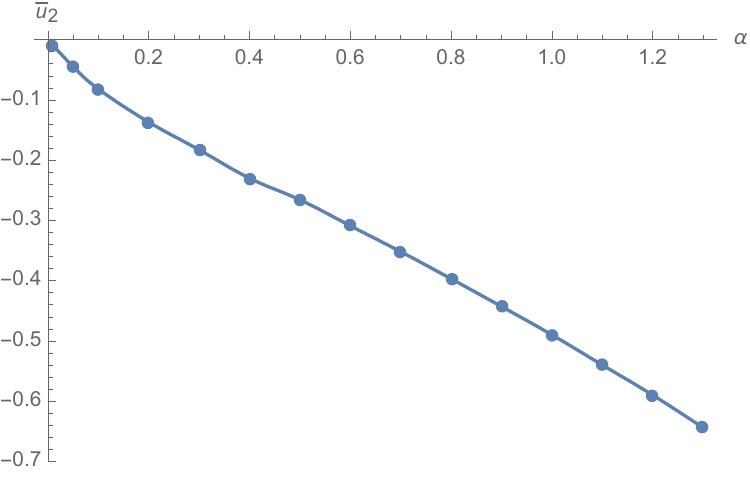}
\end{center}
\caption{Behavior of the couplings with respect to the parameter $\alpha$.}\label{depalpha2}
\end{figure}

At the order $6$, we find spurious solutions, violating the bound \eqref{boundeta}, but three physically relevant fixed point solutions: 
\begin{align}
\text{Fp1}_{61}&=\{\bar{u}_2\approx -0.49,\bar{u}_4\approx 1.58,\bar{u}_6\approx 4.18\}\\
\text{Fp2}_{61}&=\{\bar{u}_2\approx  1.13,\bar{u}_4\approx -43.35,\bar{u}_6\approx  1130.56\}\\
\text{Fp3}_{61}&=\{\bar{u}_2\approx 8.70,\bar{u}_4\approx -3370.46,\bar{u}_6\approx -1.7265\times 10^6\}\,.
\end{align}
with characteristics: 
\begin{align}
\Theta_{161}&=\{-9.84,\,-1.73,\,0.78\}\,,\qquad \nu_{161}\approx +0.29\,,\\
\Theta_{261}&=\{2.38,\,1.57,-0.17\}\,,\qquad \quad\nu_{261}\approx -0.30\,,\\
\Theta_{361}&=\{-1.70,\,-0.30,\,1.64\}\,,\qquad \nu_{361}\approx -0.73\,,
\end{align}
The last fixed point has a strong but negative sextic coupling and poses a stability issue. Interestingly however two fixed points with the same characteristics are recovered at order $8$, and one of them, whose characteristics are close to $\text{Fp1}_{61}$ is recovered also at order $10$ and beyond, seeming to improve its reliability, and then:
\begin{claim}
We conjecture the existence of an IR critical fixed point with a single relevant direction.\label{claimUVFP2}
\end{claim}
Note that the value of the critical exponents along the single irrelevant direction is almost the same with truncations of order $6$ ($\theta=0.780$), $8$ ($\theta=0.745$) and $10$ ($\theta=0.739$). The robustness of this result can be improved again by investigating the dependency on the regulator. Indeed, even if it is difficult to investigate the full dependency of the truncation on the regulator, it is suitable to investigate the dependency of the critical exponent on a few numbers of parameters, defining a restricted family of regulators. For instance, we can consider the one-parameter family:
\begin{equation}
R_k^{(\alpha)}(p_1,p_2):=\alpha Z(k)k \left(2-\frac{p_1+p_2}{k}\right) f\left(\frac{p_1}{k}\right)f\left(\frac{p_2}{k}\right)\,,
\end{equation}
and to evaluate the dependency of critical exponents on the value of the parameter $\alpha$ \cite{canet2003optimization,duclut2017frequency}. The dependency on $\alpha$ of the relevant critical exponent $\theta$ and of the anomalous dimension are shown in Figure \ref{depalpha}. It is difficult to identify a region where the dependency over $\alpha$ vanishes, as in the previous references. However, one can notice that, around $\alpha=0$, the dependency is small enough, and furthermore, $\theta$ seems to reach a finite value:
\begin{equation}
\theta(\alpha \to 0) \to 0.50\,.
\end{equation}
Note that, even though the anomalous dimension seems to vanish (as the quartic coupling) in this limit, this is not the case for the sextic coupling, which reaches a finite and positive value (see Figure \ref{depalpha2}). Then, the fixed point does not collapse over the Gaussian one and corresponds again to an interacting fixed point in the vanishing regulator limit. Such a limit has furthermore been considered recently in the literature -- see \cite{baldazzi2021limit}, and seems to be also relevant in that case. 

\begin{figure}
\begin{center}
\includegraphics[scale=0.5]{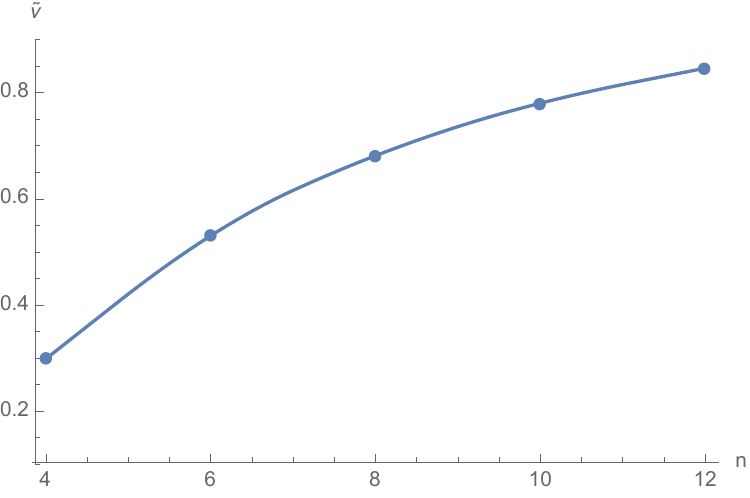}\quad \includegraphics[scale=0.5]{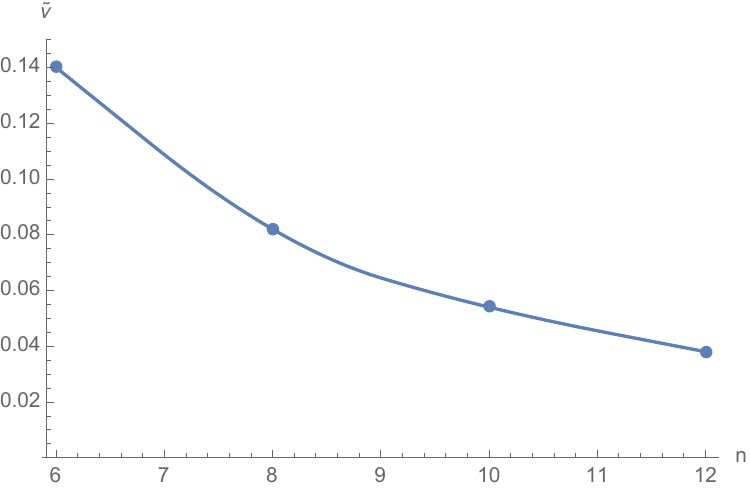}
\end{center}
\caption{Behavior of the anomalous dimension with the order of the truncation for the purely attractive fixed point (on the left) and the purely repulsive fixed point (on the right).}\label{figbehaviornu}
\end{figure}

\paragraph{Deep IR, scaling 2.} In the deep IR limit, and if $\sigma=\mathcal{O}(1)$, $\tilde{\sigma}$ must be arbitrarily large, and we will investigate the limit $\tilde{\sigma}\to \infty$ of the flow equations. At the quartic order, we get only a single physically relevant fixed point, 
\begin{equation}
\mathcal{F}p1_{4}\approx (\tilde{u}_2=-0.5,\, \tilde{u}_4 \approx 1.56 \,\tilde{\sigma}^{3/2})\,,
\end{equation}
which has only irrelevant directions:
\begin{equation}
\Theta_{\mathcal{F}41}\approx (-2.21,\,-0.92)\,,\qquad \tilde{\nu}_{41}\approx 0.30\,.
\end{equation}
A fixed point with the same characteristics is recovered for higher truncations (we checked it until order 20), in all cases, we find purely attractive fixed points, with large critical exponents, and increasing anomalous dimension ($\tilde{\nu} \sim 0.8$ at order $10$ -- see Figure \ref{figbehaviornu}). Then, because the fixed point survives around higher-order truncations and plays in favor of its reliability, the fact that the anomalous dimension becomes large should be the consequence of a breakdown of the derivative expansion. A consistency check could be to investigate contributions beyond the LPA, but this goes out from the scoop of this paper and will be the subject of \cite{LahocheSamaryPrepa}, in preparation. Then at this stage and in the absence of additional arguments, we conjecture that

\begin{claim}
It must exist a purely attractive global IR fixed point (or more precisely asymptotically in a fixed trajectory). 
\end{claim}

Note that, another reason why we can doubt the existence of such a fixed point comes from the power counting: because all the quartic interactions are relevant, the derivative quartic interactions are in principle irrelevant. However, these effects could compete with higher-order interactions, for larger truncations.
\medskip

In addition, until order $6$, a fixed point appears, which is purely repulsive toward IR scales, and for coupling values:
\begin{equation}
\mathcal{F}p2_{6}\approx (\tilde{u}_2=-0.28,\, \tilde{u}_4 \approx 1.68 \,\tilde{\sigma}^{3/2},\, \tilde{u}_6 \approx -9.50 \,\tilde{\sigma}^{3})\,,
\end{equation}
with critical exponents and anomalous dimension:
\begin{equation}
\Theta_{\mathcal{F}p2_6}\approx (0.34,\,0.71,\,0.81)\,,\qquad \tilde{\nu}_{61}\approx 0.14\,.
\end{equation}
Even though the sign of the larger coupling changes with the truncation, making the stability difficult to ensure, this fixed point (which survives on to the higher order truncations) seems to be more reliable than the previous one (see Figure \ref{figbehaviornu}). We thus complete the previous claim with the following:

\begin{claim}
It must exist as a purely repulsive global IR fixed point (or more precisely, the asymptotic of a fixed trajectory). 
\end{claim}

\begin{figure}[H]
\begin{center}
\includegraphics[scale=0.4]{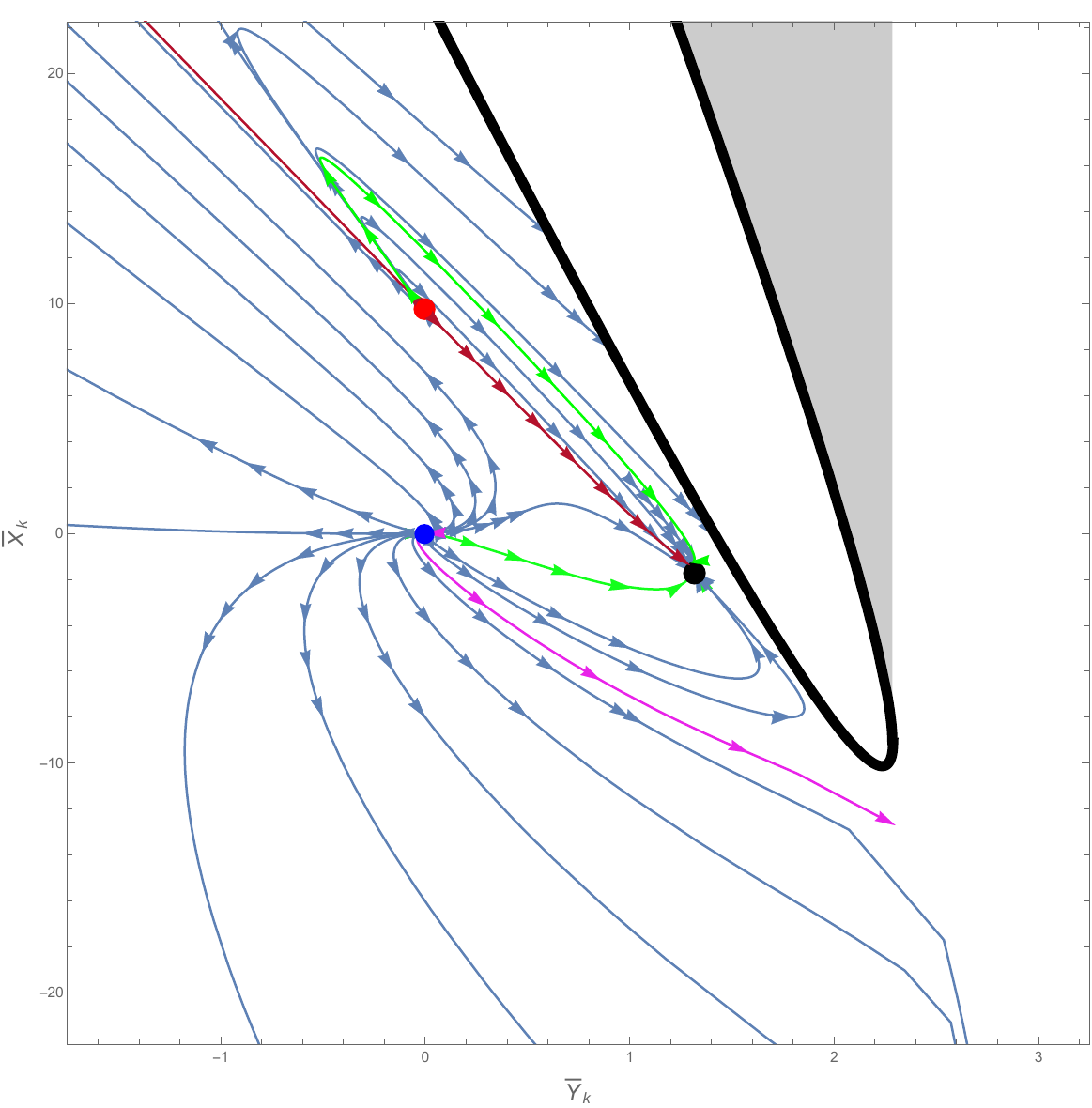}\includegraphics[scale=0.7]{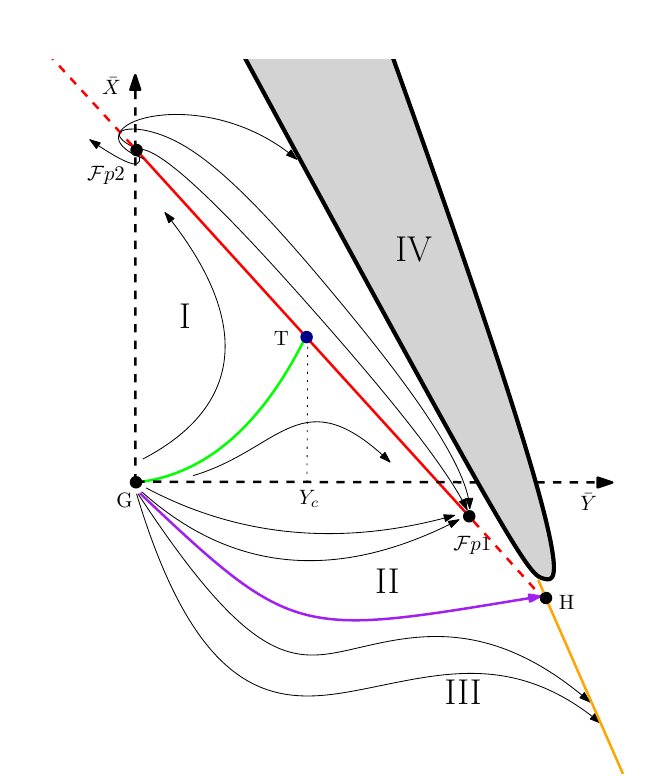}
\end{center}
\caption{On the left: Behavior of the renormalization flow in the plane spanned by the two interacting fixed points and the Gaussian one. On the right: Qualitative illustration of the different phases. In the region in the interior of the thick black line, the denominator of $\tilde{\nu}$ is negative.}\label{figflowIR}
\end{figure}

The conclusions are summarized in Figure \ref{figflowIR}, which shows the behavior of the renormalization group flow for the sextic truncation, in the plane spanned by the three fixed points $\mathcal{F}p1_{6}$, $\mathcal{F}p2_{6}$ and $G$, the Gaussian fixed points. The coordinates on the plan are denoted as $\bar{X}$ and $\bar{Y}$, and the plot is for the remaining coordinate $\bar{Z}=0$. Explicitly\footnote{\textbf{Nota}: we rescaled the coupling to cancel the dependency on $\tilde{\sigma}$ of the couplings at the fixed point.}:
\begin{align}
\bar{X}&:=-0.03\tilde{u}_2+0.17 \tilde{u}_4-0.98 \tilde{u}_6\\
\bar{Y}&:=-0.59 \tilde{u}_2 + 0.80 \tilde{u}_4 +0.16 \tilde{u}_6\\
\bar{Z}&:=0.81 \tilde{u}_2 + 0.58 \tilde{u}_4 + 0.08 \tilde{u}_6
\end{align}
The structure of the numerical phase diagram on the left of Figure \ref{figflowIR} is reproduced qualitatively on the figure on the right. Let us comment on the structure of the diagram. First, there are three distinguished points:
\begin{enumerate}
    \item The three fixed points $\mathcal{F}p1_{6}$, $\mathcal{F}p2_{6}$ and $G$.
    \item The two points $T$ and $H$, which are endpoints of some critical lines, separate two distinct regimes (as the green and the purple lines do in the figure on the right). 
\end{enumerate}
 There are in addition some relevant lines (Figure \ref{figflowIR} on the right)
 \begin{enumerate}
     \item The red line, connecting the two interacting fixed points.
     \item The green line between G and T, separating regions I and II.
     \item The portion of the green line from T to $\mathcal{F}p2$ which is a \textit{discontinuity line}.
     \item The purple line, separates regions II and III.
     \item The solid black line, where the denominator of the anomalous dimension vanishes. 
 \end{enumerate}
 and some relevant regions:
 \begin{enumerate}
     \item Region I, where the flow lines escape from the Gaussian region with $\bar{X}>0$ and $\bar{Y}<0$.
     \item Region II where the flow ends on $\mathcal{F}p1$
     \item Region III where the flow reaches the orange line.
     \item Region IV, where the denominator of the anomalous dimension is negative.
 \end{enumerate}

 \begin{remark}
It is important to notice that the flow diagram in Figure \ref{figflowIR} can be viewed as a snapshot of the stream, because of the $\tilde{\sigma}$ dependency. 
 \end{remark}

\begin{figure}[H]
\begin{center}
\includegraphics[scale=1]{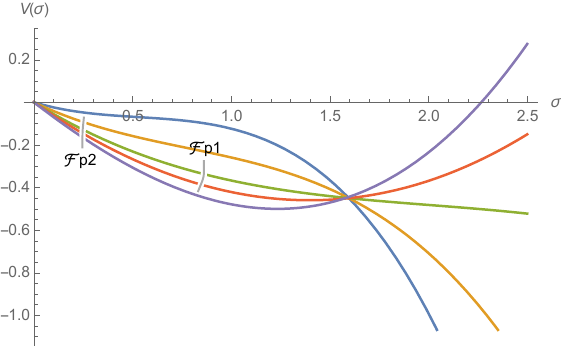}
\end{center}
\caption{On the left: Asymptotic behavior of the effective potential along the red line, between interacting fixed points.}\label{figflowIR01}
\end{figure}

Let us investigate the behavior of the renormalization group flow between the different regions of the phase diagram of Figure \ref{figflowIR}. Between $\mathcal{F}p2$ and $\mathcal{F}p1$, along the red curve crossing the point T, with the equation:
\begin{equation}
\bar{X}\approx 9.7\, -8.9 \bar{Y}\,.
\end{equation}
Usually, the different phases of a field theory are distinguished by the value taken by some parameter called \textbf{order parameter}, like the magnetization for magnetic systems of the density for fluids, conductivity for superconductors, and so on \cite{Zinn-Justin:1989rgp}. Indeed, the precise nature of the order parameter depends on the problem and the nature of the transition we consider. Here, it is difficult to consider usual indicators, like magnetization, which is nothing but the mean field, because the non-local nature of the interaction makes the vacuum solution for the potential difficult to read. For this reason, we propose to investigate the characteristics of the different phases from the large $\N$ spectrum of the IR effective complex matrix theory, such that kinetic effects can be neglected. This is the best compromise regarding the non-locality of the interaction, and similar to what we do for instance in magnetic systems, where we characterize phase for the minima of the effective potential. A method to compute the effective eigenvalue distribution from the potential of a complex random matrix is recalled in Appendix \ref{App6}, and we will apply this method along the red line connecting the two interacting fixed points. More precisely, we consider the asymptotic eigenvalue distribution $\mu(\lambda)$, where $\lambda$ are the eigenvalues of the \textit{positive} hermitian matrix:
\begin{equation}
\chi:=M^\dagger M\,.
\end{equation}
The effective potential, in the sextic truncation, is shown on the left of Figure \ref{figflowIR01}, where $\sigma$ denote here some eigenvalue for $\chi$, and the shape of the potential show that eigenvalues could be confined around $0$, and we can consider a solution for the effective distribution $\mu(\lambda)$, bounded between $a\geq 0$ and some positive number $b>a$ at the tail of the spectrum we have to determine (see  \eqref{formulamu} for more detail). First, we set $a=0$ and show that this is the only physically relevant condition in this region of the phase space.
\medskip

There are essentially two conditions we impose for $b$: 1) the distribution has to vanish for $\lambda \to b$ and the distribution has to be normalizable. These two conditions allow fixing both $b$ and the integration constant $C$ in the integral formula \eqref{formulamu}. The normalization constraint takes the form of a condition $\mathrm{Sol}(b)=0$, where:
\begin{equation}
\mathrm{Sol}(b):=\int_0^b \mu_b(\lambda) \extd \lambda-1\,,
\end{equation}
$\mu_b$ being the solution for the eigenvalue distribution that vanish for $\lambda=b$ but for $b$ unfixed. 

\begin{figure}[H]
\begin{center}
\includegraphics[scale=0.73]{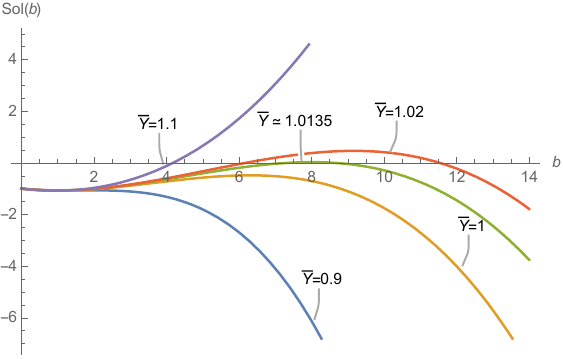}\quad \includegraphics[scale=0.73]{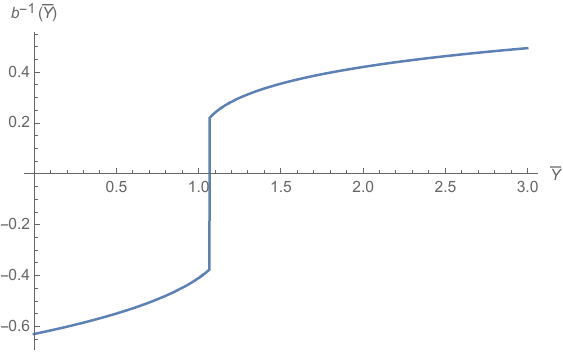}
\end{center}
\caption{On the left, the behavior of the function $\mathrm{Sol}(b)$, for $\bar{Y}$ fixed. On the right: Behavior of the inverse of the solution for the edge boundary with $\bar{Y}$, along the red curve. $\bar{Y}_c \approx 1.0135$.}\label{figflowIR01Prime}
\end{figure}

On the left of Figure \ref{figflowIR01Prime}, we show the behavior of the function $\mathrm{Sol}(b)$ along the red line, for different values for $\bar{Y}$. We distinguish clearly two phases:
\begin{itemize}
    \item For $\bar{Y}<\bar{Y}_c$, no solutions are found, the potential is not deep enough and the effective distribution is unbounded. Accordingly, with the standard one-dimensional Coulomb Gas analogy, this corresponds to a dilute phase.

    \item For $\bar{Y}\geq \bar{Y}_c$, a solution is found (only the first zero is relevant), corresponding to a condensed phase. 
\end{itemize}

\begin{figure}[H]
\begin{center}
\includegraphics[scale=0.8]{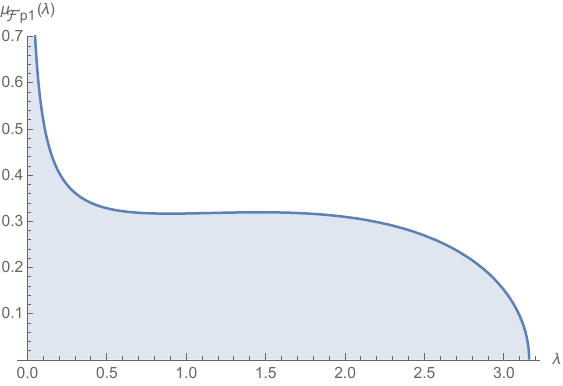}\includegraphics[scale=0.8]{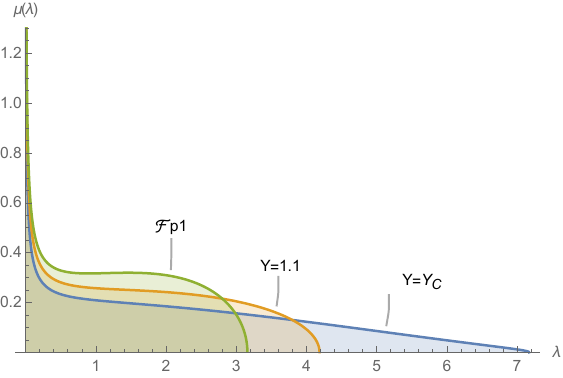}
\end{center}
\caption{On the left: Behavior of the asymptotic distribution at the fixed point $\mathcal{F}p1$. On the right: Evolution of the effective eigenvalue distribution along the red curve.}\label{figflowIR001}
\end{figure}

The critical value $\bar{Y}_c$ is estimated for $\bar{Y}_c \approx 1.0135$ (it corresponds to the abscissa coordinate of the point T, see Figure \ref{figflowIR}), and the shape of the eigenvalue distribution from $Y_c$ to $\mathcal{F}p1$ is shown on Figure \ref{figflowIR001}. Note that the system transits spontaneously from a phase where $b^{-1}=0$ to a phase where, $b^{-1}\neq 0$ at the transition point $\bar{Y}_c$, a discontinuity reminiscent of the \textit{first order phase transition} \cite{van1993regularity}. On the right of Figure \ref{figflowIR01Prime}, we show the behavior of the function $b^{-1}(\bar{Y})$, the dependency of the solution of the equation $\Sol(b)=0$ with $\bar{Y}$. 

\begin{figure}[H]
\begin{center}
\includegraphics[scale=0.8]{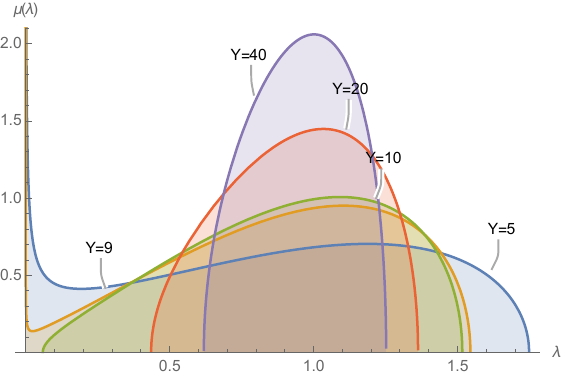}\includegraphics[scale=0.8]{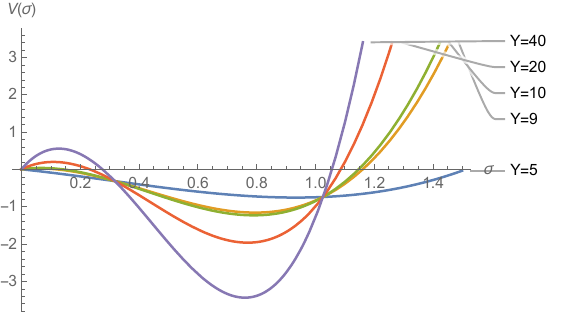}
\end{center}
\caption{On the left: Behavior of the effective IR eigenvalue distribution along the critical line. On the right: Effective potential along the critical curve.}\label{figflowIR2}
\end{figure}
Now, let us consider the orange curve, from point H in Figure \ref{figflowIR}, with the equation:
\begin{equation}
\bar{X}=+ 9.65\, -8.87\, \bar{Y}\,.
\end{equation}

The evolution of the potential and the effective distribution is shown in Figure \ref{figflowIR3}. As $\bar{Y}$ increases, the deep of the well increases, until the repulsive effect between the eigenvalues is no longer sufficient to allow the distribution to spread out along the y-axis at the origin. Note that, the distribution stops abruptly for a certain value of $a\neq 0$ (see equation \eqref{formulamu}). Indeed, at the transition point, the solution assumed to start at the origin (i.e. for $a=0$) becomes negative around the origin, whereas at the same time, the solution for $a\neq 0$ becomes relevant (see Figure \ref{figflowIR3}). The transition point is numerically evaluated at the value:
\begin{equation}
\bar{\mathcal{Y}}_{c}\approx 9.45\,.
\end{equation}
Asymptotically, for $\bar{Y}$ large enough, the effective distribution is more and more peaked around the value $\lambda=1$, with vanishing variance (see Figure \ref{figflowIR3}). 

\begin{figure}[H]
\begin{center}
\includegraphics[scale=0.8]{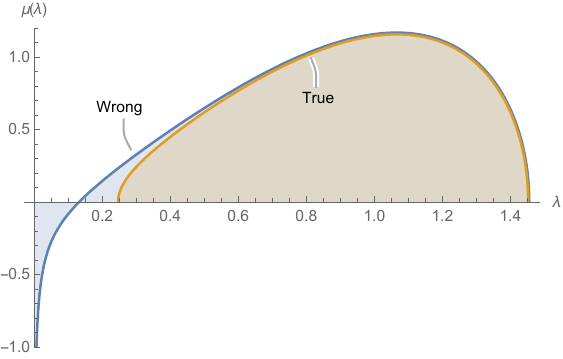}\includegraphics[scale=0.8]{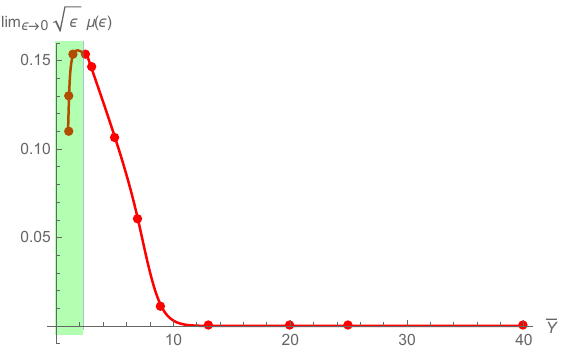}
\end{center}
\caption{On the left: Behavior of the wrong and true distribution near the transition point for $\bar{Y}=13$. On the right: Behavior of the distribution at the edge during transition. In the green region, we follow the red line since point H, where we start to follow the orange line.}\label{figflowIR3}
\end{figure}

\begin{figure}[H]
\begin{center}
\includegraphics[scale=0.73]{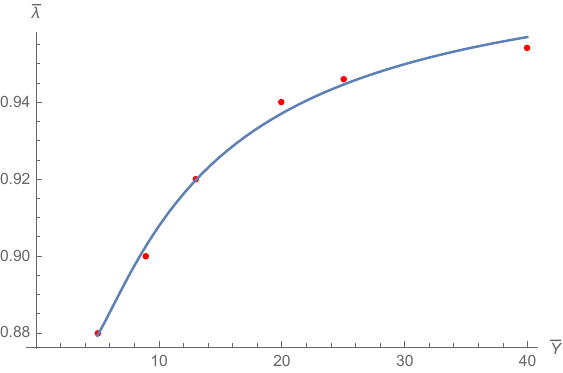}\qquad\includegraphics[scale=0.73]{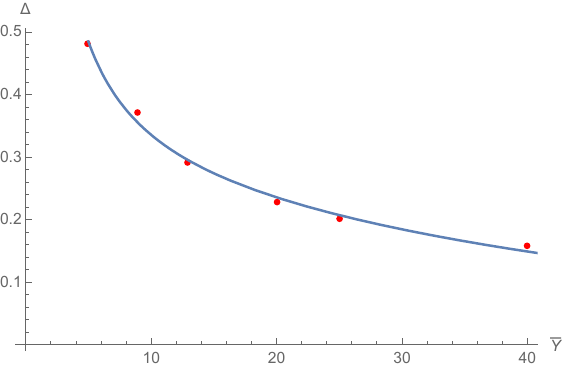}
\end{center}
\caption{Behavior of the average eigenvalue and the standard deviation along the critical line.}\label{figflowIR3}
\end{figure}

In contrast with what we observed previously, the transition seems to be continuous: the value of $a$ passes continuously from $a=0$ to $a\neq 0$, and the characteristics of the distributions seem to be continuous as well at the transition point, as we can see for instance on the Figure \ref{figflowIR3}. These characteristics refer to a \textit{second order phase transition.}

\subsection{Comments about the different scaling laws}

In the previous subsection, we investigated the behavior of the renormalization group flow in different regimes, with different choices of regularizations and different scaling. This also reflects a different point of view on the equilibrium distribution. One can indeed view it 
\begin{enumerate}
    \item As the equilibrium distribution corresponding to the maximum entropy states of the Langevin equation \eqref{eq1},
    \item Intrinsically, as the Boltzmann distribution of a specific random matrix model. 
\end{enumerate}
The two points of view are not equivalents.  In the first case, time provides an intrinsic notion of scale, whereas in the second point of view, no intrinsic notion of scale exists \textit{à priori}. This is reminiscent of what happens in tensorial field theories, where no structured background space-time exists, but an intrinsic notion of dimension emerges from the way the Feynman amplitudes diverges \cite{Lahoche:2018oeo}. Here, no divergence is expected, and because no intrinsic dimension is attached to the integral over $p$ (which is the limit of a sum), we are free to consider the most appropriate scaling depending on the regime we focus on. In contrast, if we choose the first interpretation, an intrinsic notion of dimension exists because time provides an extra notion of dimension. Then in that case, if $k^{-1}$ is a timescale, the dimension of couplings is fixed from the requirement that MSR action is dimensionless, and (as we will see in the next section), we recover exactly the scaling law \eqref{scalinlaw2} in that case. Hence, taking into account the extra-time dimension, and choosing $k^{-1}$ as a timescale, is equivalent to the renormalization group scheme pictured in Figure \ref{figactivepassive}, where the standard deviation $\sigma$ has dimension $1$ and is then re-scaled at each renormalization group step.

\section{Conclusion and outlooks}
\label{Conclusion}

We have developed the nonperturbative renormalization group techniques for the equilibrium states of a specific kind of stochastic matrices, characterized by a disorder coupling materialized by a Wigner matrix. In the large $\N$ limit, this disorder looks like an effective kinetics, whose eigenvalues are distributed accordingly with the Wigner semi-circle.  These specific kinetics and the nonlocal nature of the interaction attribute original properties to the renormalization group, and in particular, the canonical dimensions, instead of being fixed, acquire a non-trivial dependence on the scale. Note that this non-conventional fact will be observed also recently in a series of papers \cite{LahocheSignal2022,lahoche2021generalized,2021,lahoche2021signal} (by the same authors) aiming to use a renormalization group for signal detection in nearly continuous spectra. This dependence notably renders obsolete the notion of fixed point, so dear to the study of the ordinary renormalization group. In this article, we considered notions allowing us to get closer, to those of fixed trajectories and asymptotic fixed points, allowing us to discuss the existence of stable fixed structures in the IR regime, where the scale dependency of the canonical dimension is small enough, and close to the power counting of a three-dimensional quantum field theory. 
\medskip

We constructed and investigated different scaling and regularization schemes, providing a general formalism to construct the renormalization group, for different approximation procedures. Focusing on the derivative and vertex expansions, our main statement is the existence of a reliable enough first-order phase transition between a dilute high-temperature phase and a condensed phase. From a dynamical point of view, this transition is interpreted as the transition from an equilibrium phase to a phase where the system never reaches equilibrium, the formal equilibrium state then being non-integrable (see \eqref{normalizationcondition}). This statement seems however difficult to confirm at this stage because the sign of the larger coupling at the fixed point oscillates. 
\medskip

These difficulties, we expect, arise because the approximation we considered in this paper has to be considered incomplete for a fundamental reason, and we have a solid reason to doubt our large-order truncations. Indeed, in contrast with ordinary local quantum field theory, the anomalous dimension is nonzero also in the symmetric phase, and we neglected the momenta dependence of the effective vertices to construct our approximation. We expect these effects could be in competition with higher-order couplings, for large truncation and must be investigated carefully. This will be the topic of the companion paper \cite{LahocheSamaryPrepa}, aiming to construct an advanced approximation scheme in the deep UV, closing the flow equation hierarchy using the nontrivial Ward identities arising because of the effective kinetics. 
\medskip

\bigskip 


\bigskip

\appendix
\begin{center}
\begin{LARGE}
\textbf{Additional material}
\end{LARGE}
\end{center}

\section{Fokker-Planck equation for complex matrices}\label{App1}

In this appendix, we propose for the reader unfamiliar with the stochastic formalism to derive the Fokker-Planck equation for the complex matrix (more detail about the general formalism can be found in the standard reference \cite{ZinnJustinBook2,Zinn-Justin:1989rgp}). We denote as $\textbf{q}:=M(t)$, $\textbf{q}_0:=M(t=0)$, and $\bm{q}_x:= \{ \RE(M)_{ij}\}$ and  $\bm{q}_y:= \{ \IM(M)_{ij}\}$
 the $N^2$ vectors corresponding to the real and imaginary parts of the matrix $M$. In the same way, we denote as $B_x$ and $B_y$ the real and imaginary parts of the noise matrix field, and as $\textbf{f}_{ix}$ and $\textbf{f}_{iy}$ the real and imaginary parts of $\textbf{f}_i$, the $i$th component of the $2\N^2$ vector:
\begin{equation}
\textbf{f}=\left\{\RE \left(\frac{\delta \mathcal{H}}{\delta M_{ij}}\right)\,;\IM\left(\frac{\delta \mathcal{H}}{\delta M_{ij}}\right) \right\}\,.\label{expfi}
\end{equation}
The discrete probability $W(\epsilon,\textbf{q},\textbf{q}_0):= P[\textbf{q},t_0+\epsilon; \textbf{q}_0,t_0]$ to jump from $\textbf{q}_0:=M(t=0)$ to $\textbf{q}:=M(t)$ in the laps $\epsilon$ is then:
\begin{align}
W(\epsilon,\textbf{q},\textbf{q}_0)&=\frac{1}{z_\eta} \exp{\left(-\frac{(\textbf{q}-\textbf{q}_0+\textbf{f}_i \epsilon)^2_{x}}{\epsilon T}-\frac{(\textbf{q}-\textbf{q}_0+\textbf{f}_i \epsilon)^2_y}{\epsilon T}\right)}\,,
\end{align}
and the Fourier transform concerning the variable $\textbf{q}$ reads:
\begin{equation}
W(\epsilon,\textbf{q},\textbf{q}_0)=e^{-i(\textbf{q}_{0x}\cdot \textbf{p}_{x}+\textbf{q}_{0y}\cdot \textbf{p}_{y})} \left(1-\frac{\epsilon T}{4} (\textbf{p}^2_{x}+\textbf{p}^2_y)+i( \textbf{f}_{ix}\cdot \textbf{p}_x+ \textbf{f}_{iy}\cdot \textbf{p}_y) \epsilon\right) +\mathcal{O}(\epsilon^2)
\end{equation}
Taking the inverse Fourier transform, the Hamiltonian then reads:
\begin{equation}
\textbf{H}=\sum_{\alpha} \frac{\partial}{\partial x_\alpha} \left(\frac{T}{4}\frac{\partial}{\partial x_\alpha}+ f_{\alpha i x} \right)+\sum_{\alpha} \frac{\partial}{\partial y_\alpha} \left(\frac{T}{4}\frac{\partial}{\partial y_\alpha}+ f_{\alpha i y} \right)
\end{equation}
Using the fact that $\frac{\partial}{\partial \bar{z}}=\frac{1}{2}\left(\frac{\partial}{\partial x}+i \frac{\partial}{\partial y}\right)$, we have for instance:
\begin{equation}
f_{\alpha i x}:=\RE \left(\frac{\partial H^{(\mathbb{C})}}{\partial M_{ij}}\right) \Big\vert_{\alpha\equiv (i,j)} = \frac{1}{2} \frac{\partial H^{(\mathbb{C})}}{\partial x_{\alpha}},
\end{equation}
then
\begin{equation}
\textbf{H}=\sum_{\alpha} \frac{\partial}{\partial x_\alpha} \left(\frac{T}{4}\frac{\partial}{\partial x_\alpha}+ \frac{1}{2} \frac{\partial H^{(\mathbb{C})}}{\partial x_{\alpha }} \right)+\sum_{\alpha} \frac{\partial}{\partial y_\alpha} \left(\frac{T}{4}\frac{\partial}{\partial y_\alpha}+\frac{1}{2} \frac{\partial H^{(\mathbb{C})}}{\partial y_{\alpha}} \right)\,.
\end{equation}
We also have the relation:
\begin{equation}
\frac{T}{4} \sum_{\alpha} \left( \frac{\partial^2}{\partial x_\alpha^2}+\frac{\partial^2}{\partial y_\alpha^2}\right)= T\sum_{i,j} \frac{\partial^2}{\partial M_{ij}\partial \bar{M}_{ij}}\,.
\end{equation}
Furthermore,  for some function $\mathcal{F}$ we get the relation
\begin{align}
\frac{\partial}{\partial \bar{z}} \left(\frac{\partial H^{(\mathbb{C})}}{\partial z}\right)\mathcal{F}= \frac{\partial^2 H^{(\mathbb{C})}}{\partial \bar{z}\partial z}\mathcal{F}+\frac{\partial H^{(\mathbb{C})}}{\partial z}\frac{\partial \mathcal{F}}{\partial \bar{z}}\,,
\end{align}
and:
\begin{align}
\frac{\partial H^{(\mathbb{C})}}{\partial z}\frac{\partial \mathcal{F}}{\partial \bar{z}}&= \frac{1}{4}\left(\frac{\partial H^{(\mathbb{C})}}{\partial x}-i\frac{\partial H^{(\mathbb{C})}}{\partial y}\right)\left(\frac{\partial \mathcal{F}}{\partial x}+i\frac{\partial \mathcal{F}}{\partial y}\right)\\
&=\frac{1}{4}\left(\frac{\partial H^{(\mathbb{C})}}{\partial x}\frac{\partial \mathcal{F}}{\partial x}+\frac{\partial H^{(\mathbb{C})}}{\partial y}\frac{\partial \mathcal{F}}{\partial y}+i\frac{\partial H^{(\mathbb{C})}}{\partial x}\frac{\partial \mathcal{F}}{\partial y}-i\frac{\partial H^{(\mathbb{C})}}{\partial y}\frac{\partial \mathcal{F}}{\partial x}\right)\,,
\end{align}
leading to:
\begin{equation}
\frac{\partial H^{(\mathbb{C})}}{\partial z}\frac{\partial \mathcal{F}}{\partial \bar{z}}+\frac{\partial H^{(\mathbb{C})}}{\partial \bar{z}}\frac{\partial \mathcal{F}}{\partial {z}}=\frac{1}{2}\left(\frac{\partial H^{(\mathbb{C})}}{\partial x}\frac{\partial \mathcal{F}}{\partial x}+\frac{\partial H^{(\mathbb{C})}}{\partial y}\frac{\partial \mathcal{F}}{\partial y}\right)\,.
\end{equation}
Finally:
\begin{align}
\textbf{H}=\sum_{i,j} \left(T \frac{\partial^2}{\partial M_{ij}\partial \bar{M}_{ij}}+2\frac{\partial^2H^{(\mathbb{C})}}{\partial M_{ij}\partial \bar{M}_{ij}}+\frac{\partial H^{(\mathbb{C})} }{\partial{M_{ij}}}\frac{\partial}{\partial {\bar{M}_{ij}}}+\frac{\partial H^{(\mathbb{C})} }{\partial{\bar{M}_{ij}}}\frac{\partial}{\partial {{M}_{ij}}}\right)\,.
\end{align}

\section{One loop computation for the quartic complex matrix model}\label{App3}

In this section, we consider a perturbative point of view and compute vertex corrections and $\beta$-functions at the leading order in the coupling constant. Note that we focus on the quartic model, such that only $a_2$ is non-vanishing in the expansion \eqref{H0complex}.

\subsection{One loop vertex expansion}

Assuming we included the regulator $R_k$ in the definition of the effective bare propagator, its diagonal part reads:
\begin{equation}
G_k^{(0)}(p_1,p_2)=\frac{1}{p_1+p_2+m+R_{k}(p_1,p_2)}\,.
\end{equation}
Where the subscript ‘‘$0$'' means order zero in the loop expansion. The effective propagator, in contrast, must-reads:
\begin{equation}
G_k(p_1,p_2)=\frac{1}{p_1+p_2+m+R_{k}(p_1,p_2)-\Sigma(p_1,p_2)}\label{effectivepropaDyson}\,.
\end{equation}
where $\Sigma(p_1,p_2)$ is the \textit{self energy}, including loop effects. In this section, we only consider the Litim regulator \eqref{eqR}, which, assuming to focus on the deep IR, reads:
\begin{equation}
R_k(p,p^\prime)\approx Z(k)(2k-p-p^\prime) \theta(k-p)\theta(k-p^\prime)\,.
\end{equation}
The factor $Z(k)$, the wave function renormalization can be also computed perturbatively, and we define, at order $1$ in $a_2$:
\begin{equation}
Z(k)=1-Z^{(1)}(k) a_2+\mathcal{O}(a_2^2)\,.
\end{equation}
Hence, the presence of the factor $Z$ has to be taken into account in the expansion of $G_k$, at the same time as the one-loop self-energy $\Sigma$. Graphically, the expansion of the Dyson equation reads:
\begin{equation}
\underbrace{\vcenter{\hbox{\includegraphics[scale=0.7]{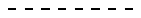}}}}_{G_k}=\underbrace{\vcenter{\hbox{\includegraphics[scale=0.7]{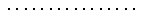}}}}_{\mathcal{O}(a_2^0)}+\underbrace{\vcenter{\hbox{\includegraphics[scale=0.7]{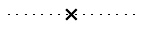}}}+\vcenter{\hbox{\includegraphics[scale=0.7]{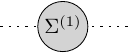}}}}_{\mathcal{O}(a_2)}+\mathcal{O}(a^2_2)\,.
\end{equation}
The rule is the following: the dashed edge is the effective propagator $G_k$, the dotted edge the bare propagator with $Z=1$, the cross means the insertion of the wave function correction $Z^{(1)}(k) a_2$ and the gray circle the insertion of the one-loop self-energy. 
\medskip

The one-loop effective $4$-point function can be computed in the same way, using a suitable graphical notation to materialize effective vertices:
\begin{equation}
\vcenter{\hbox{\includegraphics[scale=0.8]{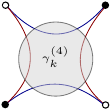}}}=\underbrace{\vcenter{\hbox{\includegraphics[scale=0.7]{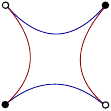}}}}_{\mathcal{O}(a_2)}\,\,+\,\,\underbrace{\vcenter{\hbox{\includegraphics[scale=0.6]{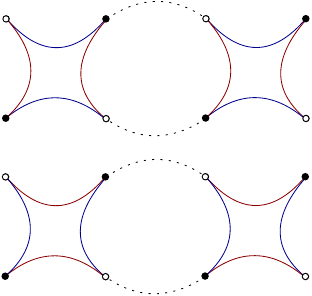}}}}_{\mathcal{O}(a_2^2)}\,\,+\,\,\mathcal{O}(a_2^3)\,.\label{4pointseries}
\end{equation}
where the function $\gamma_k^{(4)}:\mathbb{R}^4\to \mathbb{R}$ depends only on the four real external momenta components, because of the momenta conservation assumed for external faces (see definition \ref{defface}). 
\medskip

Now, let us compute each term. First, the one-loop self-energy comes from the diagram (barred dotted edges mean the edge is amputated):
\begin{equation}
\vcenter{\hbox{\includegraphics[scale=1]{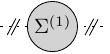}}}=\vcenter{\hbox{\includegraphics[scale=1]{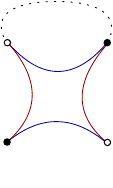}}}\,+\,\vcenter{\hbox{\includegraphics[scale=1]{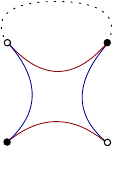}}}\,.\label{sum1loop}
\end{equation}
Explicitly, and taking into account symmetry factors, we have:
\begin{align}
\vcenter{\hbox{\includegraphics[scale=1]{selfE1.pdf}}}&=-2 \left( \frac{a_2}{4 \N}\right) \sum_{p'}\frac{1}{m+p+p'+R_k(p,p^\prime)} \\
&\rightarrow -\frac{a_2}{2}\int_0^{4\sigma}\,\frac{\sqrt{p^\prime(4\sigma-p^\prime)}}{2\pi\sigma^2}\frac{\extd p^\prime}{m(p)+p^\prime+R_k(p,p^\prime)}\,,
\end{align}
where $p$ denote the external momenta running through the loop, $p^\prime$ is the momenta around the internal face, and:
\begin{equation}
m(p):=m+p\,.
\end{equation}
The previous integral splits into two contributions, depending on the momenta along the loop is larger or smaller than $k$:
\begin{equation}
\int_0^{4\sigma}\,\frac{\sqrt{p^\prime(4\sigma-p^\prime)}}{2\pi\sigma^2}\frac{\extd p^\prime}{m(p)+p^\prime+R_k(p,p^\prime)}=I_1(k,p)+I_2(k,p)\,,
\end{equation}
where:
\begin{equation}
I_1(k,p):=\int_k^{4\sigma}\,\frac{\sqrt{p^\prime(4\sigma-p^\prime)}}{2\pi\sigma^2}\frac{dp^\prime}{m(p)+p^\prime}
\end{equation}
and
\begin{align}
\nonumber I_2(k,p)&:=\int_0^{k}\,\frac{\sqrt{p^\prime(4\sigma-p^\prime)}}{2\pi\sigma^2}\frac{\extd p^\prime}{m(p)+p^\prime+(2k-p-p^\prime)\theta(k-p)}\\\nonumber
&=\frac{1}{2k+m}\int_0^{k}\,dp^\prime\frac{\sqrt{p^\prime(4\sigma-p^\prime)}}{2\pi\sigma^2}\theta(k-p)\\
&\,\,+\int_0^{k}\,\frac{\sqrt{p^\prime(4\sigma-p^\prime)}}{2\pi\sigma^2}\frac{dp^\prime}{m(p)+p^\prime}\theta(p-k)\,.
\end{align}
Explicitly, assuming $m>0$ and $k\ll 1$, we get, setting $\sigma=1$:
\begin{equation}
I_1(k,p):=\frac{1}{2} \left[-\frac{4 \left(\sqrt{k}-\sqrt{m(p)} \tan ^{-1}\left(\sqrt{\frac{k}{m(p)}}\right)\right)}{\pi }+m(p)-\sqrt{m(p) (m(p)+4)}+2\right] \label{eqI1}
\end{equation}
and
\begin{equation}
I_2(k,p):=\frac{2 k^{3/2}}{3 \pi(2k+m(p))}\theta(k-p)+\frac{2 \left(\sqrt{k}-\sqrt{m(p)} \tan ^{-1}\left(\sqrt{\frac{k}{m(p)}}\right)\right)}{\pi }\theta(p-k)\,.\label{eqI2}
\end{equation}
Because of the sum \eqref{sum1loop}, the one-loop self energy splits in two terms,
\begin{equation}
\Sigma(p_1,p_2)=\sigma(p_1)+\sigma(p_2)\,,
\end{equation}
where:
\begin{equation}
\boxed{
\sigma(p):=-\frac{a_2}{2}\,\left(I_1(k,p)+I_2(k,p)\right).}\label{formulaSigma}
\end{equation}

We proceed on the same way for the $4$-point diagram, that contribute to the 1PI $4$-points function $\gamma^{(4)}_k(p_1,p_2,p_3,p_4)$ in \eqref{4pointseries}. The one loop contribution being of order $a_2^2$, it makes sense to set $Z(k)=1$ during calculation. By direct inspection, the one-loop contribution has the following structure:
\begin{equation}
\gamma^{(4)}_k(p_1,p_2,p_3,p_4)=-\frac{a^2_2}{\N}\left(g_k^{(1)}(p_1,p_2)+g_k^{(1)}(p_3,p_4)\right)\,,
\end{equation}
the two terms of order $a_2^2$ in \eqref{4pointseries}, and the one-loop kernel $g_k^{(1)}(p_1,p_2)$ can be computed explicitly using Feynman rules as:
\begin{align}
&g_k^{(1)}(p_1,p_2):= \frac{1}{\N}\sum_{p}\frac{1}{(m+p_1+p+R_k(p,p_1))(m+p_2+p+R_k(p,p_2))}\cr
&\rightarrow  \int_0^{4\sigma}\,\frac{\sqrt{p(4\sigma-p)}}{2\pi\sigma^2}\frac{\extd p}{(m(p_1)+p+R_k(p_1,p))(m(p_2)+p+R_k(p_2,p))}\,,
\end{align}
and can be computed, assuming $m>0$, as:
\begin{align}
\nonumber g_k^{(1)}(p_1,p_2)&:=J_{1,k}(p_1,p_2)\theta(p_1-k)\theta(p_2-k)+J_{2,k}(p_1,p_2)\theta(k-p_1)\theta(p_2-k)\\
&+J_{3,k}(p_1,p_2)\theta(p_1-k)\theta(k-p_2)+J_{4,k}(p_1,p_2)\theta(k-p_1)\theta(k-p_2)\,,\label{formulaGamma4}
\end{align}
where, explicitly assuming $k$ is small enough:
\begin{align}
&\nonumber J_{1,k}(p_1,p_2):= \frac{p_2-p_1+\sqrt{m\left(p_1\right) \left(m\left(p_1\right)+4\right)}-\sqrt{m\left(p_2\right) \left(m\left(p_2\right)+4\right)}}{2 \left(p_1-p_2\right)}
\end{align}
\begin{align}
& J_{2,k}(p_1,p_2):=-\frac{2 \sqrt{m\left(p_1\right)} \tan ^{-1}\left(\sqrt{\frac{k}{m\left(p_1\right)}}\right)-2 \sqrt{m\left(p_2\right)} \tan ^{-1}\left(\sqrt{\frac{k}{m\left(p_2\right)}}\right)}{\pi  m\left(p_1\right)-\pi  m\left(p_2\right)}\\\nonumber
&+\frac{p_2-p_1+\sqrt{m\left(p_1\right) \left(m\left(p_1\right)+4\right)}-\sqrt{m\left(p_2\right) \left(m\left(p_2\right)+4\right)}}{2 \left(p_1-p_2\right)}\\
&+\frac{2 \left(\sqrt{k}-\sqrt{m\left(p_2\right)} \tan ^{-1}\left(\sqrt{\frac{k}{m\left(p_2\right)}}\right)\right)}{\pi  \left(2 k+m\left(p_1\right)\right)}
\end{align}
\begin{align}
&\nonumber J_{4,k}(p_1,p_2):=-\frac{2 \sqrt{m\left(p_1\right)} \tan ^{-1}\left(\sqrt{\frac{k}{m\left(p_1\right)}}\right)-2 \sqrt{m\left(p_2\right)} \tan ^{-1}\left(\sqrt{\frac{k}{m\left(p_2\right)}}\right)}{\pi  m\left(p_1\right)-\pi  m\left(p_2\right)}\\\nonumber
&+\frac{p_2-p_1+\sqrt{m\left(p_1\right) \left(m\left(p_1\right)+4\right)}-\sqrt{m\left(p_2\right) \left(m\left(p_2\right)+4\right)}}{2 \left(p_1-p_2\right)}\\
&+\frac{2 k^{3/2}}{3 \pi  \left(2 k+m\left(p_1\right)\right) \left(2 k+m\left(p_2\right)\right)}\,,
\end{align}
and $J_{3,k}(p_1,p_2)=J_{2,k}(p_2,p_1)$. To be complete, we have to compute $Z^{(1)}(k)$, which is nothing but:
\begin{equation}
Z^{(1)}(k)a_2=\frac{\partial}{\partial p} \sigma(p) \bigg\vert_{p=0}\,,
\end{equation}
and explicitly, assuming $k\neq 0$:
\begin{align}
\nonumber Z^{(1)}(k)=\frac{1}{4}\Bigg(\frac{2 \sqrt{k}}{\pi  k+\pi  m}&+\frac{4 \sqrt{k}}{2 \pi  k+\pi  m}-\frac{4 \sqrt{k} (5 k+3 m)}{3 \pi  (2 k+m)^2}\\
&-\frac{2 \tan ^{-1}\left(\sqrt{\frac{k}{m}}\right)}{\pi  \sqrt{m}}+\frac{m+2}{\sqrt{m (m+4)}}-1\Bigg)\,.\label{oneloopZ}
\end{align}

\subsection{Critical $\beta$-function}

In the previous subsection, we computed the vertex correction at the lading order in $a_2$. Note that, due to the non-local nature of field theory, the wave function renormalization receives a correction at the order $a_2$, when it is for order $a_2^2$ for a local field theory \cite{peskin2018introduction}. We focus on the critical regime, $0<m\ll 1$. 
\medskip

With zero external momenta, the $4$-points function reads:
\begin{equation}
\gamma_{k,0000}^{(4)}=\frac{a_2}{\N}-\frac{2a_2^2}{\N}g_k^{(1)}(0,0)\,,
\end{equation}
and we define the \textit{effective coupling} $u_4$ as:
\begin{equation}
u_4(k):=\N \gamma_{k,0000}^{(4)}\,,
\end{equation}
then, in the leading order, we have:
\begin{align}
k \frac{\extd u_4}{\extd k}&= -2a_2^2 k \frac{\extd}{\extd k}J_{4,k}^{(1)}(0,0)+\mathcal{O}(a_2^3)\\
&\approx -2u_4^2 k \frac{\extd}{\extd k}J_{4,k}^{(1)}(0,0)+\mathcal{O}(a_2^3)\,.
\end{align}
Using the explicit expression \eqref{formulaGamma4}, we have, assuming again $m>0$:
\begin{align}
k \frac{\extd}{\extd k}J_{4,k}^{(1)}(0,0)& = -\frac{k^{5/2} \left(26 k^2+37 k m+14 m^2\right)}{3 \pi  (k+m)^2 (2 k+m)^3}\\
&\approx -\frac{13}{12 \pi  \sqrt{k}}+\frac{9 m}{4 \pi  k^{3/2}}+\mathcal{O}(m^2)
\end{align}
The anomalous dimension can be also computed easily from \eqref{oneloopZ},
\begin{align}
\eta:=\frac{1}{Z(k)}k \frac{\extd}{\extd k} Z(k)&= -a_2\frac{k^{5/2} \left(26 k^2+37 k m+14 m^2\right)}{6 \pi  (k+m)^2 (2 k+m)^3} +\mathcal{O}(a_2^2)\\
&=a_2\left(-\frac{9 m}{8 \pi  k^{3/2}}+\frac{13}{24 \pi  \sqrt{k}}\right)+\mathcal{O}(a_2^2,m^2)\,,
\end{align}
and we define the renormalized and dimensionless coupling constant as:
\begin{equation}
{u}_4=:Z^2(k) k^{1/2}\tilde{u}_4\,,
\end{equation}
then, at all orders in $m$
\begin{equation}
\boxed{\frac{\extd \tilde{u}_4}{\extd t}=-\frac{1}{2} \tilde{u}_4+\frac{13\tilde{u}_4^2}{12\pi}\,,}\label{beta1}
\end{equation}
with $t:=\ln(k)$. Now, let us move on to the flow equation for mass. At one loop, the effective mass is\footnote{It can be reads directly on the expression \eqref{effectivepropaDyson}.}:
\begin{equation}
u_2(k):=m-\Sigma(0,0)=m+a_2(I_1(k,0)+I_2(k,0))\,,
\end{equation}
and, then:
\begin{align}
k\frac{\extd u_2}{\extd k}&=a_2k\frac{\extd}{\extd k}(I_1(k,0)+I_2(k,0))+\mathcal{O}(a_2^2)\\
&\approx u_4 k\frac{\extd}{\extd k}(I_1(k,0)+I_2(k,0))+\mathcal{O}(a_2^2)
\end{align}
Explicitly, we find from the expressions \eqref{eqI1} and \eqref{eqI2}:
\begin{equation}
k\frac{\extd}{\extd k}(I_1(k,0)+I_2(k,0))=\frac{-k^{5/2} (10 k+7 m)}{3 \pi  (k+m) (2 k+m)^2}=-\frac{5 \sqrt{k}}{6 \pi }+\frac{13 m}{12 \pi\sqrt{k}}+\mathcal{O}\left(m^2\right)\,,
\end{equation}
and finally:
\begin{equation}
\boxed{\frac{\extd \tilde{u}_2}{\extd t}=-\tilde{u}_2-\frac{11 \tilde{u}_4 }{8\pi}\,,}\label{beta2}
\end{equation}
where: $u_2=:Z(k)k \tilde{u}_2$. The $\beta$-functions \eqref{beta1} and \eqref{beta2} admit a Wilson-Fisher fixed point solution for the values:
\begin{equation}
\tilde{u}_4^*=\frac{6 \pi }{13}\approx 1.45\,,\qquad \tilde{u}_2^*=-\frac{33}{52}\approx -0.63\,,
\end{equation}
with critical exponents:
\begin{equation}
\theta_1=1\,,\qquad \theta_2=-\frac{1}{2}\,.
\end{equation}
Characteristics that are reminiscent of the fixed point solution \eqref{fixedpointsolC1} for the quartic truncation. 

\begin{remark}
The one-loop $\beta$-function we computed in this appendix differs from the perturbative expansion of the flow equations obtained from vertex expansion (equations \eqref{beta1Scaling1} and \eqref{beta1Scaling2}). Indeed, we get:
\begin{equation}
\frac{\extd \tilde{u}_4}{\extd t}=-\frac{1}{2} \tilde{u}_4+\frac{\tilde{u}_4^2}{3\pi}+\mathcal{O}(\tilde{u}_4)\,,\label{beta1pert}
\end{equation}
and
\begin{equation}
\frac{\extd \tilde{u}_2}{\extd t}=-\tilde{u}_2-\frac{7 \tilde{u}_4 }{24\pi}+\mathcal{O}(\tilde{u}_4)\,.
\end{equation}
The reason why the one-loop expressions (if they are correct) are different remains mysterious at this stage. It is known that they are universal for dimensionless couplings until two loops \cite{benedetti2016functional,codello2014scheme}, but the couplings we consider here being not dimensionless, the statement does not hold and could explain this disagreement, what we will investigate further. 
\end{remark}

\section{Computing eigenvalue spectra of $M^\dagger M$.}\label{App6}

\begin{figure}[H]
\begin{center}
\includegraphics[scale=1]{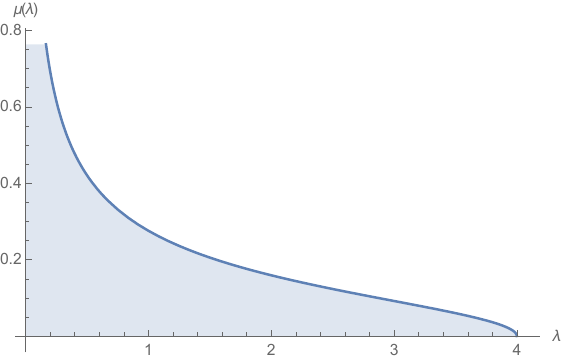}
\end{center}
\caption{Typical shape of the eigenvalue distribution with $\mu_1=1$.}\label{FigMP}
\end{figure}

In this section, we shortly review the solution of the problem to determine the eigenvalue spectra of $\chi=M^\dagger M$, in the large $\N$ limit, from the knowledge of the distribution law for the complex matrix $M$. We assume the partition is of the form:
\begin{equation}
Z:=\int \extd M \extd \bar{M} \, e^{-\mu_1 \Tr M^\dagger M - \Tr\, V(M,M^\dagger)}\,,
\end{equation}
where $V$ expands in power of $\chi$:
\begin{equation}
V(M,M^\dagger)=\sum_{p=1}^K \frac{\mu_p}{(p!)^2 \N^{p-1}}\, (M^\dagger M)^{p}\,.
\end{equation}
To perform the change of variable $\chi=M^\dagger M$ in the path integral, we use the clever expression of the identity:
\begin{equation}
1\equiv \int \extd \chi \,\delta\left(\chi-M^\dagger M\right)\,,
\end{equation}
where the integral is performed on the space of Hermitian matrices. The Dirac delta function can be furthermore rewritten in the Fourier representation for the $N(N+1)$ independent components of $\chi$:
\begin{equation}
\delta(\chi-M^\dagger M) \propto \int \extd \lambda \, e^{i\, \Tr \lambda (\chi-M^\dagger M)}\,,
\end{equation}
where the integration is performed on the hermitian matrix $\lambda$\footnote{It is suitable to add a small imaginary part proportional to the identity to make the integral well defined.}; furthermore, it is suitable to exploit the unitary invariance of the integral over $M$ to choose $\chi$ diagonal. Finally, the partition function reads: 
\begin{equation}
Z=\int \extd M \extd \bar{M} \extd \chi \extd \lambda \, e^{i\, \Tr \lambda (\chi-M^\dagger M)} e^{-\mu_1 \Tr \chi - \Tr\, V(\chi)}\,.
\end{equation}
The Gaussian integral over $M$ can be easily performed, 
\begin{equation}
\int \extd M \extd \bar{M}e^{-i\, \Tr \lambda M^\dagger M} \propto (\det \lambda)^{-N}\equiv  e^{-\N\, \Tr\,\ln (\lambda)}\,,
\end{equation}
such that the partition function becomes,
\begin{equation}
Z=\int \extd \chi \extd \lambda \,e^{-\N\, \Tr\,\ln(\lambda)-\mu_1 \Tr \chi - \Tr\, V(\chi)} e^{i\, \Tr\, \lambda \chi}\,.
\end{equation}
Let us make the change of variable $\lambda \to \sigma:= \chi^{\frac{1}{2}} \lambda \chi^{\frac{1}{2}}$, the Jacobian of the transformation is:
\begin{align}
\prod_i \extd\lambda_{ii} \prod_{j>i} \extd \RE \lambda_{ij} \extd \IM\lambda_{ij}&= \prod_{i} \chi_{ii}^{-1} \prod_{j>i} (\chi_{ii}^2\chi_{jj}^2)^{-\frac{1}{2}}\prod_i \extd\sigma_{ii} \prod_{j>i} \extd \RE \sigma_{ij} \extd \IM\sigma_{ij}\\
& =(\det \, \chi)^{-\N} \, \prod_i \extd\sigma_{ii} \prod_{j>i} \extd \RE \sigma_{ij} \extd \IM\sigma_{ij}\,.
\end{align}
Furthermore:
\begin{equation}
(\det \lambda)^{-N}=(\det (\chi^{-1} \sigma))^{-N}=(\det \sigma)^{-N} (\det \chi)^{N}\,,
\end{equation}
and:
\begin{equation}
Z\propto \int \extd \chi \,e^{-\mu_1 \Tr\, \chi - \Tr\, V(\chi)}\,.
\end{equation}
In the large $\N$ limit, the integral is dominated by the saddle point. Using the polar decomposition of the integration measure over Hermitian matrices \cite{potters2020first,Francesco_1995}:
\begin{equation}
\extd \chi= \extd \Lambda (\Delta(\Lambda))^2 \extd U\,,
\end{equation}
where $\Lambda$ is a diagonal and positive definite matrix with eigenvalues $\lambda_i$: $\extd \Lambda \equiv \prod_i \extd \lambda_i$, $\Delta(\Lambda):=\prod_{j>i} (\lambda_j-\lambda_i)$ is the Vandermonde determinant and $\extd U$ is the Haar measure over the Unitary group. It is furthermore suitable to rescale $\chi \to \N \chi$, such that the saddle point equation reads:
\begin{equation}
\frac{1}{2}\left(\mu_1 + \sum_{p=1}^K \frac{p \mu_p}{(p!)^2}\, \lambda_i^{p-1}\right)=\frac{1}{N}\sum_{j\neq i}\, \frac{1}{\lambda_i-\lambda_j}\,.
\end{equation}
This equation can be investigated in the continuum limit from Tricomi's formula \cite{tricomi1985integral,potters2020first,landau1986theory}. Indeed, denoting as $\mu(\lambda)$ the large $\N$ eigenvalues distribution, the previous equation reads:
\begin{equation}
\underbrace{\frac{1}{2}\left(\mu_1 + \sum_{p=1}^K \frac{p \mu_p}{(p!)^2}\, \lambda^{p-1}\right)}_{f(\lambda)}=\dashint\, \extd\lambda^\prime\frac{\mu(\lambda^\prime)}{\lambda-\lambda^\prime}\,.
\end{equation}
If the spectrum is bounded, the solution for $\mu(\lambda)$ as the integral form:
\begin{equation}
\mu(\lambda):=-\frac{1}{\pi^2\sqrt{(\lambda-a)(b-\lambda)}}\left(\dashint\, \extd\lambda^\prime \sqrt{(\lambda^\prime-a)(b-\lambda^\prime)} \frac{f(\lambda^\prime)}{\lambda-\lambda^\prime}+C\right)\,,\label{formulamu}
\end{equation}
where $C$ is adjusted from the boundary and normalization conditions. Consider for instance a purely quadratic confining potential, for $f(\lambda)=\mu_1/2$, we have, choosing $C$ such that $\mu(b)=0$,
\begin{equation}
\mu(\lambda)=\frac{\mu_1\sqrt{\lambda (b-\lambda)}}{2 \pi \lambda}\,.
\end{equation}
Note that it is suitable to set $a=0$, and the boundary $b$ is then fixed by the normalization condition:
\begin{equation}
\int_0^b \extd \lambda \mu(\lambda)=1 \quad \to \quad b=\frac{4}{\mu_1}\,,
\end{equation}
and finally:
\begin{equation}
\mu(\lambda)=\frac{\mu_1 \sqrt{\lambda\left(\frac{4}{\mu_1}-\lambda\right)}}{2 \pi  \lambda}\,.
\end{equation}
Not surprisingly, the distribution looks like a Marchenko–Pastur distribution and the typical shape is shown in Figure \ref{FigMP}. 

The formula \eqref{formulamu} can be generalized in cases where we have only one hard wall or no hard wall. In the first case, assuming for instance we have a hard wall at $x=b$, the solution is \cite{landau1986theory}:
\begin{equation}
\mu(\lambda):=-\frac{1}{\pi^2} \sqrt{\frac{\lambda-a}{b-\lambda}}\dashint\, \extd\lambda^\prime \sqrt{\frac{b-\lambda^\prime}{\lambda^\prime-a}} \,\frac{f(\lambda^\prime)}{\lambda-\lambda^\prime}\,,
\end{equation}
and in the second case:
\begin{equation}
\mu(\lambda):=-\frac{1}{\pi^2} \sqrt{(\lambda-a)(b-\lambda)}\dashint\, \extd\lambda^\prime \frac{1}{\sqrt{(\lambda^\prime-a)(b-\lambda^\prime)}}\, \frac{f(\lambda^\prime)}{\lambda-\lambda^\prime}\,.
\end{equation}

\printbibliography[heading=bibintoc]
\end{document}